\documentclass[fleqn,usenatbib]{mnras}

\usepackage{newtxtext,newtxmath}

\usepackage[T1]{fontenc}
\usepackage{ae,aecompl}

\usepackage{graphicx}	

\usepackage{soul} 
\usepackage{xargs} 
\usepackage[pdftex,dvipsnames]{xcolor}  
\usepackage{amsmath}
\usepackage{rotating}
\usepackage{pdflscape}
\definecolor{shade}{gray}{0.9}

\newcommand{\hst }{{\em HST}}
\newcommand{\gaia}{{\em Gaia}}
\newcommand{\cxo }{{\em CXO}}

\title[Where are the magnetar binaries?]{Where are the magnetar binary companions? Candidates from a comparison with binary population synthesis predictions}

\author[A. A. Chrimes et al.]{A. A. Chrimes,$^{1}$\thanks{E-mail: a.chrimes@astro.ru.nl}
A. J. Levan$^{1,2}$,
A. S. Fruchter$^{3}$,
P. J. Groot$^{1,4,5}$,
P. G. Jonker$^{1,6}$,
C. Kouveliotou$^{7,8}$,\newauthor
J. D. Lyman$^{2}$,
E. R. Stanway$^{2}$,
N. R. Tanvir$^{9}$
and K. Wiersema$^{10}$\\
$^{1}$Department of Astrophysics/IMAPP, Radboud University, P.O. Box 9010, 6500 GL, The Netherlands \\
$^{2}$Department of Physics, University of Warwick, Gibbet Hill Road, Coventry, UK \\
$^{3}$Space Telescope Science Institute, 3700 San Martin Drive, Baltimore, MD 21218, USA \\
$^{4}$Inter-University Institute for Data Intensive Astronomy, Department of Astronomy, University of Cape Town, Private Bag X3, Rondebosch 7701, South Africa \\
$^{5}$South African Astronomical Observatory, P.O. Box 9, 7935 Observatory, South Africa \\
$^{6}$SRON, Netherlands Institute for Space Research, Niels Bohrweg 4, 2333 CA, Leiden, The Netherlands \\
$^{7}$Department of Physics, The George Washington University, Corcoran Hall, 725 21st St NW, Washington, DC 20052, USA \\
$^{8}$GWU/Astronomy, Physics and Statistics Institute of Sciences (APSIS) \\
$^{9}$School of Physics and Astronomy, University of Leicester, University Road, Leicester, LE1 7RH, UK \\
$^{10}$Physics Department, Lancaster University, Lancaster, LA1 4YB, UK \\
}

\date{Accepted XXX. Received YYY; in original form ZZZ}

\pubyear{2022}

\begin{document}
\label{firstpage}
\pagerange{\pageref{firstpage}--\pageref{lastpage}}
\maketitle

\begin{abstract}
It is well established that magnetars are neutron stars with extreme magnetic fields and young ages, but the evolutionary pathways to their creation are still uncertain. Since most massive stars are in binaries, if magnetars are a frequent result of core-collapse supernovae, some fraction are expected to have a bound companion at the time of observation. In this paper, we utilise literature constraints, including deep Hubble Space Telescope imaging, to search for bound stellar companions to magnetars. The magnitude and colour measurements are interpreted in the context of binary population synthesis predictions. We find two candidates for stellar companions associated with CXOU\,J171405.7--381031 and SGR\,0755--2933, based on their $J$-$H$ colours and $H$-band absolute magnitudes. Overall, the proportion of the Galactic magnetar population with a plausibly stellar near-infrared counterpart candidate, based on their magnitudes and colours, is between 5 and 10 per cent. This is consistent with a population synthesis prediction of 5 per cent, for the fraction of core-collapse neutron stars arising from primaries which remain bound to their companion after the supernova. These results are therefore consistent with magnetars being drawn in an unbiased way from the natal core-collapse neutron star population, but some contribution from alternative progenitor channels cannot be ruled out.
\end{abstract}




\section{Introduction}
Magnetars comprise a class of highly magnetised young neutron stars, with magnetic field strengths up to 10$^{15}$\,G \citep{1998Natur.393..235K}. They are typically discovered when they produce ${\gamma}$-ray or X-ray bursts during active episodes; their nature is also confirmed through the detection of their persistent X-ray emission \citep{2017ARA&A..55..261K}. Some were originally classified as soft gamma repeaters (SGRs), others as anomalous X-ray pulsars (AXPs), but both populations are now believed to be manifestations of the same phenomenon \citep{1996ApJ...473..322T}. Approximately 30 are currently known; two are in the Magellanic clouds \citep[one each,][]{2014ApJS..212....6O}, while the rest reside in our Galaxy. Their very rare Giant Flares are bright enough to be seen at extragalactic distances \citep{2005Natur.434.1098H,2005Natur.434.1107P}; two originated from magnetars in the Milky Way and a handful have been identified in other galaxies \citep[e.g.][]{2010MNRAS.403..342H,2021ApJ...907L..28B}.

Measurements of magnetar spin periods $P$ and period derivatives $\dot{P}$ imply extremely young ages of 10$^{3}$-10$^{5}$\,yr, under the assumption of a surface dipole field and magnetic braking. These estimates are unreliable for magnetars, since the fundamental assumption of a constant magnetic field (and hence braking index) is shown to vary significantly \citep{2017ApJ...851...17Y}. However, these are upper limits for their decaying magnetic fields,
leading to {\it overestimates} of their true ages. The young magnetar ages are further backed up by their associations with star forming regions and supernova remnants
\citep[e.g.][]{2006ApJ...636L..41M,2007ApJ...667.1111G,2015PASJ...67....9N}. Moreover, current sophisticated methods now exist that account for their magnetic field decay, in addition to particle winds and gravitational wave emission \citep{2021ApJ...913L..12M}.

Despite some associations with H\,{\sc II} regions and supernova remnants, around half of the magnetar population have not yet been associated with a high-mass progenitor \citep{2014ApJS..212....6O}. This may be attributed to detection biases due to e.g., dust, source crowding, and high column densities. There are, however, magnetars for which there is no obvious association with recent star formation despite a relatively clear sightline (see e.g., SGR\,0501$+$4516 which lies in the Galactic anti-centre direction). Putative non-core-collapse production channels include binary neutron star mergers and the accretion or merger induced collapse of ONe white dwarfs resulting in magnetars \citep{1992ApJ...392L...9D,2006MNRAS.368L...1L,2021arXiv211012140A}.

The rarity of magnetars \citep{1994Natur.368..125K,2019MNRAS.487.1426B} raises questions about their origin. Are specific conditions required to generate neutron stars with such strong magnetic fields? If their magnetic fields decay fast enough, it may be that a large fraction of core-collapse supernovae result in magnetars \citep{2015PASJ...67....9N,2019MNRAS.487.1426B}, which then may decay into X-ray dim isolated neutron stars \citep[XDINS,][]{2021MNRAS.tmp.2446J} before fading beyond detection limits. Correspondingly, if magnetars are intrinsically rare, particular evolutionary pathways may be required. Suggestions include core-collapse following main sequence mergers \citep{2020MNRAS.495.2796S}, rapid core rotation pre-collapse \citep{2021arXiv211101814W}, and non-core collapse channels, as mentioned above.

Owing to their large energy reservoirs, magnetars have also been invoked as the central engines in a variety of transients, from compact object mergers, which produce short gamma-ray bursts \citep[GRBs,][]{2012MNRAS.419.1537B,2013MNRAS.431.1745G,2013MNRAS.430.1061R,2014MNRAS.438..240G,2014MNRAS.439.3916M,2021MNRAS.508.2505Z}, core-collapse long GRBs \citep{2000ApJ...537..810W,2007MNRAS.382.1029K,2011MNRAS.413.2031M}, superluminous supernovae \citep{2010ApJ...717..245K,2010ApJ...719L.204W,2015MNRAS.454.3311M} and fast blue optical transients \citep{2018ApJ...865L...3P,2020ApJ...888L..24M}. Many fast radio burst models also implicate magnetars \citep[e.g.][]{2014MNRAS.442L...9L,2019MNRAS.485.4091M}, a connection strengthened by the discovery of low-luminosity FRBs from the Galactic magnetar SGR\,1935$+$2154 \citep{2020Natur.587...54C,2020Natur.587...59B,2021NatAs...5..408Y}. The host environments of extragalactic FRBs are broadly representative of where we expect to find young neutron stars \citep{2020ApJ...903..152H,2020ApJ...905L..30S,2021ApJ...907L..31B,2021ApJ...917...75M,2021MNRAS.508.1929C,2021arXiv210801282B,2021arXiv211010847K}. However, there is also evidence that FRBs do not follow the cosmic star formation rate density \citep{2020ApJ...895L..22J,2021arXiv210907558Z}, and there are other objects which seem inconsistent with core-collapse magnetars. For example, assuming that magnetars do produce FRBs, \citet{2021ApJ...908L..12T} find that FRB\,20180916B is too offset from a nearby star forming region to be a young ejected magnetar. Subsequently, FRB\,20200120E was located in a globular cluster outside M81 \citep{2021arXiv210511445K}. Since globular clusters have old stellar populations, a young core-collapse magnetar explanation for this source is challenging \citep{2021arXiv210704059L,2021ApJ...917L..11K}. 

It remains unclear which progenitor channels contribute to magnetar production, or if any particular channel dominates their formation rate. If we assume that magnetars are drawn from the natal core-collapse neutron star population in an unbiased way, then some fraction should have a bound companion star. The majority of massive stars are in binaries \citep[e.g.][]{2014ApJS..215...15S}, and not all supernovae in binaries unbind the system, as evidenced by the existence of binary neutron star and black hole mergers, X-ray binaries and pulsars in binaries. 

In this paper, we use published photometry of counterpart candidates in conjunction with distance and extinction estimates to determine whether any of the Galactic magnetar near-infrared (NIR) counterparts could be a bound companion star. We determine the fraction of the population that could plausibly have a bound companion, and compare these results to population synthesis predictions for the fraction of neutron stars born with a bound secondary companion.

The paper begins with population synthesis predictions in Section \ref{sec:bpass}, and a exploration of the major sources of uncertainty in these predictions. We describe the sample selection in Section \ref{sec:cp}, before applying distance modulus and extinction corrections in Section \ref{sec:corr}, allowing us to compare population synthesis predictions with observations in Section \ref{sec:compare}. We summarise the results in Section \ref{sec:summary}, discuss our findings in Section \ref{sec:discuss} in the context of magnetar progenitor and evolutionary models, and conclude in Section \ref{sec:conc}. All magnitudes are quoted in the Vega system, where conversions from AB magnitudes \citep{1983ApJ...266..713O} have been made using stsynphot\footnote{\url{https://github.com/spacetelescope/stsynphot_refactor}} for Hubble Space Telescope (\hst\ ) data. AB magnitudes are obtained by adding 0.8227, 0.9204, 1.0973, and 1.2741 to Vega magnitudes for F110W, F125W, F140W and F160W respectively. For $J$, $H$ and $K$-band filters, the conversions of \cite{2007AJ....133..734B} are used.

\begin{figure}
	\includegraphics[width=0.99\columnwidth]{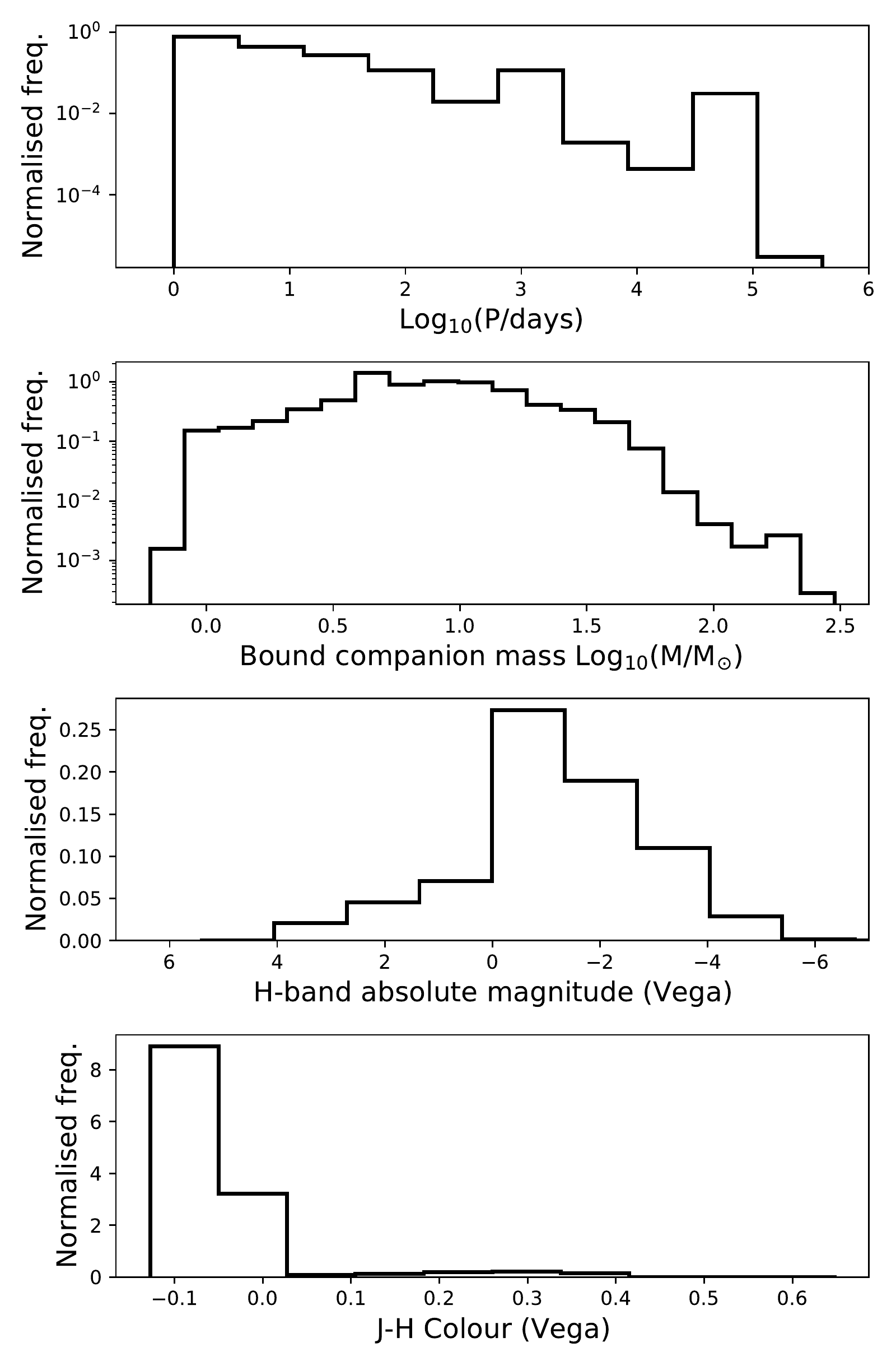}
    \caption{The predicted orbital period, mass, $H$-band absolute magnitude and $J$--$H$ colour distributions for bound companions to natal neutron stars. The median absolute $H$-band magnitude is -1.2, and the median colour is -0.08. The axes of the top two panels are on a log scale to clearly show the distribution shape across the full range of parameter space.}
    \label{fig:massH}
\end{figure}

\section{Binary Population Synthesis}\label{sec:bpass}
\subsection{Predictions}
To establish predictions for the fraction of neutron stars born in primary supernovae, and the fraction of those with a surviving bound companion, we make use of the population synthesis code BPASS \citep[Binary Population and Spectral Synthesis,][v2.2.1]{2017PASA...34...58E,2018MNRAS.479...75S}. BPASS consists of thousands of binary and single stellar evolution models across a grid of masses, mass ratios, orbital periods, and metallicities. Each model is assigned a weighting, corresponding to how often it is predicted to occur in a 10$^{6}$ M$_{\odot}$ stellar population, with the initial weightings based on observations of Galactic populations \citep[][]{2017ApJS..230...15M}. 

For a fiducial estimate, we use the default BPASS model weightings and broken power law initial mass function (IMF), with a maximum zero age main sequence mass of 300\,M$_{\odot}$, and gradients of --1.30 (0.1-0.5\,M$_{\odot}$) and --2.35 \citep[0.5-300\,M$_{\odot}$, based on][]{1993MNRAS.262..545K}. Solar metallicity (a metal mass fraction of 0.02) is assumed. To identify models which produce neutron stars in core-collapse, we require in the final time step of each model (i) a total mass $>2.0$\,M$_{\odot}$, (ii) a CO core mass $>1.38$\,M$_{\odot}$, (iii) a non-zero ONe core mass, and (iv) a remnant mass in the range $1.38<M_{\rm rem}<3.0$ \citep[see also][]{2017PASA...34...58E,2019MNRAS.482..870E,2020MNRAS.491.3479C,2021arXiv211108124B}. We then sum the weightings for all models which satisfy these criteria, and examine the contributions from different model types (single, primary, bound secondary, unbound secondary). To identify how many primary neutron stars are born with a bound companion, the weightings for secondary star models whose companion is a neutron star at $t=0$ are summed. We find that around half of all neutron stars are born in the supernovae of primary stars in binaries. Another ${\sim}$25 per cent arise from single stars or mergers, and the remaining ${\sim}$25 per cent from secondary stars. 

Around ${\sim}$10 per cent of neutron stars born from primaries remain bound after the supernova. We therefore predict that the fraction of natal neutron stars with a bound, pre-supernova companion is f$_{\rm bound}=0.05$. Figure \ref{fig:massH} shows the predicted Log$_{10}$(P/days), Log$_{10}$(M/M$_{\odot}$), $H$-band absolute magnitude and $J$--$H$ colour distributions for bound companions to neutron stars, in the first time step of secondary BPASS models. The bulk of the colours correspond to massive main sequence companions, with the smaller, redder bump arising from a minority of models where the secondary has already evolved off the main sequence when the primary goes supernova. This is in qualitative agreement with previous predictions for surviving supernova companions \citep[e.g.][]{2009ApJ...707.1578K,2017ApJ...842..125Z}.

We focus on NIR colours because of the available observational constraints on Galactic magnetar counterparts. Since the population lies predominantly in the Galactic plane, and out to large distances ($>10$\,kpc), high extinctions typically render the optical magnitudes of magnetar counterparts extremely faint (with only a few detections reported). Conversely, there $\sim$15 (depending on the confidence of the association) counterpart candidates in the $H$-band (see Section \ref{sec:cp}). Since the companions to natal neutron stars are primarily expected to be massive, un-evolved main sequence stars, other works in this area - for example searches for surviving companions to supernovae - typically use bluer optical bands \citep{2016ApJ...833..128M}. While such companions will be intrinsically fainter in the NIR, typically by $\sim$2 magnitudes compared to the optical, just $A_{\rm V}\sim2$ of extinction negates this advantage. Approximately two magnitudes of optical extinction is the lowest value inferred for any Galactic magnetar (see Section \ref{sec:corr}), and as extinction increases, the NIR becomes relatively more favourable. Combined with many more detections of magnetar counterparts in the NIR compared with the optical, this justifies our focus on the NIR as the best wavelength range in which to search for companions.

\begin{figure}
    \includegraphics[width=0.99\columnwidth]{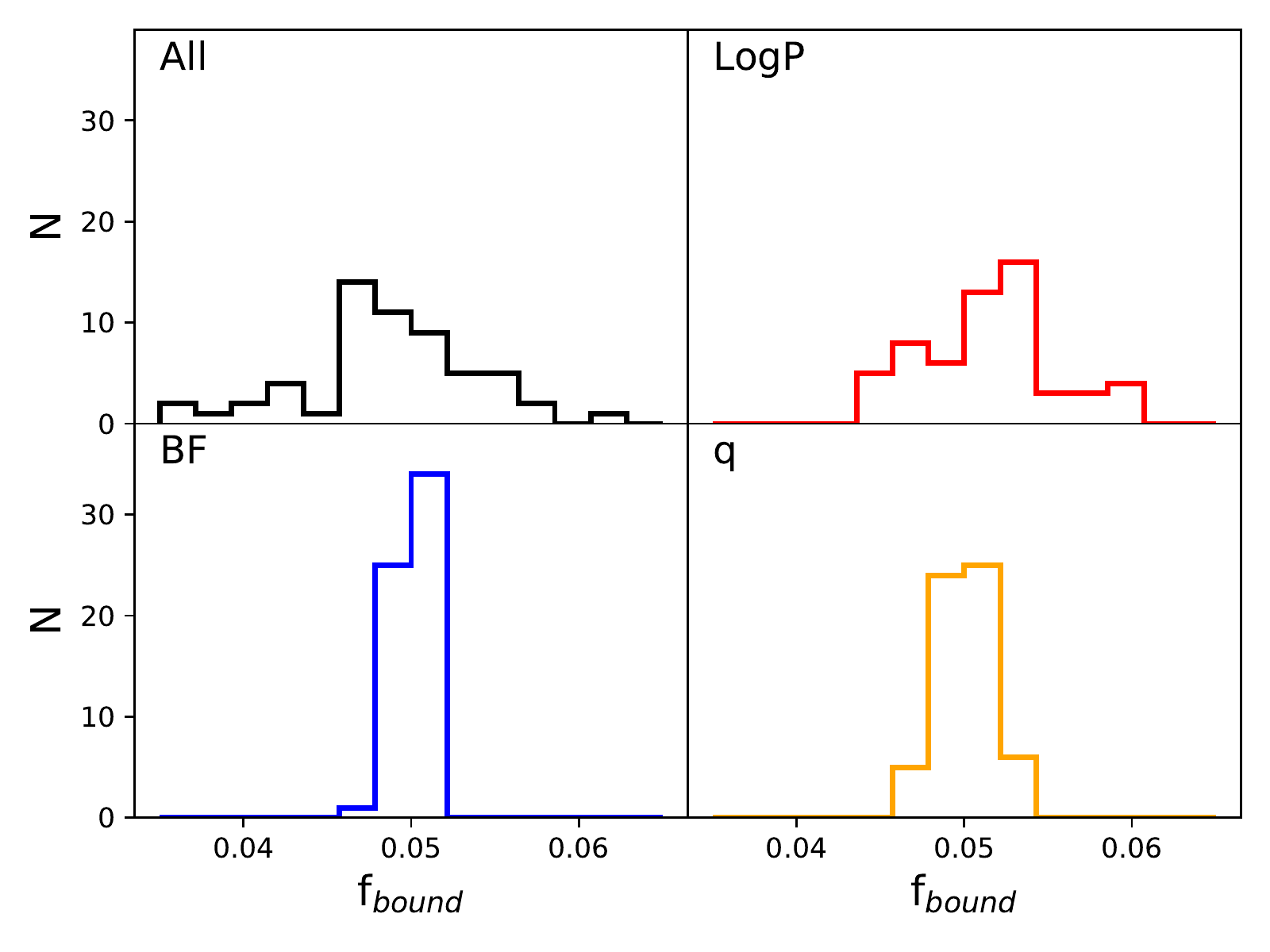}
    \caption{How the prediction for f$_{\rm bound}$ varies with 60 different realisations of the input distributions, drawn through bootstrap re-sampling \citep[as in ][]{2020MNRAS.495.4605S}. Default assumptions yield f$_{\rm bound}=0.05$. The upper left distribution arises from varying the orbital periods (Log$_{10}$(P/days), labelled LogP), mass ratios ($q$) and binary fraction (BF) simultaneously. Orbital periods are the largest individual contributor to the uncertainty.}
    \label{fig:psuncert}
\end{figure}

\subsection{Uncertainties}
Previous estimates for f$_{\rm bound}$ have varied, in part due to uncertainties on the population synthesis input distributions \citep[for details of how BPASS handles binary dynamics following a supernova, see][]{1998A&A...330.1047T,2011MNRAS.414.3501E}. For example, \citet{2011MNRAS.414.3501E} predict that the number of neutron stars born from primaries which remain bound is a few per cent. The choice of binary fraction, initial period and mass ratio distributions will change f$_{\rm bound}$, as will the assumed natal kick distribution.

\citet{2020MNRAS.495.4605S} investigated the impact of uncertainties in the BPASS input parameters on the integrated light of stellar populations. Using the same set of iterations across orbital period, mass ratio and binary fraction, and drawing from the observed distribution of these inputs in the Galactic stellar populations, we determine the impact on f$_{\rm bound}$. The default period distribution (LogP), binary fraction (BF) and mass ratio (q) distributions are based on table 13 of \citet{2017ApJS..230...15M} as follows,
\begin{enumerate}
    \item LogP: A period distribution is applied in each of the five mass bins studied by \citet{2017ApJS..230...15M}. For example, 0.095$\pm$0.018 of Solar mass stars have a companion with LogP$=$4.5--5.5 (P in days), whereas 0.30$\pm$0.09 of stars above 16\,M$_{\odot}$ have a companion in this range,
    \item BF: defined as one minus the single star fraction in each mass bin. The BF in each mass bin has an associated observational uncertainty, e.g. the single star fraction for Solar-type stars is 0.60$\pm$0.04, and among the most massive stars it is 0.06$\pm$0.06,
    \item q: the mass ratio distribution is a broken power law following $p_{q}\propto q^{\gamma}$, which varies in each LogP and (primary star) mass bin. For example, the probability $p_{q}$ of a mass ratio in the range 0.3--1.0 for Solar mass primaries in a binary with LogP$=$5 is given by $\gamma=-0.5\pm0.3$. In the mass ratio range 0.1--0.3, $\gamma=0.3\pm0.4$.
\end{enumerate}
Finally, the elevated fraction of twin systems is taken into consideration, and the distributions are interpolated across the BPASS model grid. Randomly sampling these input distributions within the quoted uncertainties \citep[assuming Gaussian uncertainties, as fully described in][]{2020MNRAS.495.4605S} yields the results shown in Figure \ref{fig:psuncert}. Uncertainty on the initial orbital periods of massive binaries is the largest source of binary parameter uncertainty in f$_{\rm bound}$. The result when all uncertainties are varied together is 0.049$\pm$0.005, the wider spread showing that there are correlated uncertainties between the parameters.

We also investigate the impact of assuming different neutron star natal kick velocities - the kick distributions of \citet{2005MNRAS.360..974H} and \citet{2020MNRAS.494.3663I} yield very similar f$_{\rm bound}$ values of 0.049 and 0.050. \citet{2021MNRAS.508.3345I} also found little difference between these kicks in terms of the Be star mass distribution in X-ray binaries. The kick of \citet{2016MNRAS.461.3747B,2018MNRAS.480.5657B}, however, gives a substantially larger value of 0.23. For an observational comparison, we refer to \cite{2021MNRAS.507.5832K} who performed a {\it Gaia} search for unbound companions to 10 supernova remnants. Combining these results with a previous search for surviving bound companions in supernova remnants, they find that 0.036--0.197 (90 per cent confidence, excluding triple systems) of core-collapse neutron stars are born in primaries which remain bound post supernova. This figure also agrees well with the results of the StarTrack binary population synthesis code \citep[yielding 0.05, assuming an 84 per cent binary fraction;][]{2019MNRAS.485.5394K} and binary$\_$c \citep[$0.11^{+0.07}_{-0.09}$, assuming 22 per cent of massive binaries merge before the first supernova,][]{2019A&A...624A..66R}. Therefore, the prediction of f$_{\rm bound}=0.05$ is consistent with other population synthesis codes and observational constraints.

\section{Sample Selection}\label{sec:cp}
Our sample consists of the 23 Galactic magnetars (excluding the Magellanic clouds) which have imaging detections or limits, and a distance estimate. Of the 31 objects listed in the McGill magnetar catalogue \citep{2014ApJS..212....6O}\footnote{\url{http://www.physics.mcgill.ca/~pulsar/magnetar/main.html}}, CXOU\,J010043.1--721134 and SGR\,0526--66 are in the Large and Small Magellanic clouds respectively, while SGRs\,1801--23 and 1808--20 have neither imaging nor a distance estimate. Swift\,J1818.0-1607 and PSR\,J1846--0258 have no photometric data while SGR\,1833--0832 and AX\,J1818.8--1559 do not have a distance estimate. The full list of magnetars in the sample, and the photometry used, is provided in Table \ref{tab:data}. Most sources have {\em HST} F160W ($H$) and F125W ($J$) observations, as reported in \citet{2022MNRAS.tmp..864C}. The NIR counterparts are localised, where possible, through astrometric alignment of a {\it Chandra} X-ray localisation with the {\em HST} images, by finding common sources in the field or doing so through an intermediate image. The RMS residuals of this translation, combined with the source cetroiding uncertainty, determines the size of the error circle in the {\em HST} frame. In other cases, a radio localisation is used (for which the absolute astrometric uncertainty of {\it HST} dominates), or simply the absolute astrometric uncertainty of the X-ray observations, if no common sources could be found and no alternative positional measurements were available. Shortened magnetar names, as indicated in this table, are used hereafter. Given the population synthesis predictions, we would expect 1$\pm$1 (Poisson uncertainty only) of the 23 objects in the sample to have a bound companion. This prediction holds under the assumptions that (i) the young age estimates are broadly correct (since the probability of a companion going supernova within the age of the magnetar is extremely low) and (ii) they are drawn from the natal core-collapse neutron star population in an unbiased away.

\begin{table*}
\centering 
\caption{The full magnetar sample considered in this paper (23 objects). Photometry (Vega magnitudes) for the most likely counterpart(s) is listed in each case. Shortened magnetar names, as in the first column without brackets, are used throughout this paper. The criteria for selection are a distance estimate and imaging. 1RXS\,J1708 has three candidate counterparts in (or just outside) the X-ray error circle \citep[see][]{2022MNRAS.tmp..864C}, the candidates A and B are labelled following \citet{2003ApJ...589L..93I} and \citet{2006ApJ...648..534D}. The magnitudes listed here are the primary ones used (i.e. for the colour-magnitude diagram), other values from the literature are also plotted later in Figure \ref{fig:bound_H}, following references in the McGill catalogue \citep{2014ApJS..212....6O}$^{2}$.}
\label{tab:data}
\begin{tabular}{llllll}
\hline %
Magnetar	&	$H$-band (Vega)	&	$J$-band (Vega)	&	Instrument/survey	&	Filter(s)	&	Reference	\\
\hline											
4U\,0142($+$61)	&	20.80$\pm$0.01	&	 21.72$\pm$0.01	&	\hst\ /WFC3	&	F160W$+$F125W	&	\citet{2022MNRAS.tmp..864C}	\\
SGR\,0418($+$5729)	&	$>$25.12	&	$>$26.08	&	\hst\ /WFC3	&	F160W$+$F125W	&	\citet{2022MNRAS.tmp..864C}	\\
SGR\,0501($+$4516)	&	22.56$^{+0.06}_{-0.07}$	&	23.33$\pm$0.07	&	\hst\ /WFC3	&	F160W$+$F125W	&	\citet{2022MNRAS.tmp..864C}	\\
1E\,1048(.1--5937)  	&	22.15${\pm}$0.20	&	 $>$24.08	&	\hst\ /NICMOS	&	F160W$+$F110W	&	\protect{\citet{2008ApJ...677..503T}}	\\
1E\,1547(.0--5408)	&	$>$20.22	&	$>$22.45	&	\hst\ /WFC3	&	F160W$+$F125W	&	\citet{2022MNRAS.tmp..864C}	\\
PSR\,J1622(--4950)	&	22.39$\pm0.05$	&	23.95$\pm0.08$	&	\hst\ /WFC3	&	F160W$+$F125W	&	\citet{2022MNRAS.tmp..864C}	\\
SGR\,1627(--41)	&	20.48$\pm$0.01	&	21.97$\pm0.01$	&	\hst\ /WFC3	&	F160W$+$F125W	&	\citet{2022MNRAS.tmp..864C}	\\
CXOU\,J1647(10.2--455216)$\dagger$	&	21.0$\pm0.1$	&	23.5$\pm0.2$	&	VLT/NACO	&	$H$,$J$	&	\protect{\cite{2018MNRAS.473.3180T}}	\\
1RXS\,J1708(49.0--400910) A	&	18.77${\pm}$0.01	&	20.61${\pm}$0.01	&	\hst\ /WFC3	&	F160W$+$F125W	&	\citet{2022MNRAS.tmp..864C}	\\
1RXS\,J1708(49.0--400910) B	&	20.58${\pm}$0.01	&	22.37$\pm$0.02	&	\hst\ /WFC3	&	F160W$+$F125W	&	\citet{2022MNRAS.tmp..864C}	\\
1RXS\,J1708(49.0--400910) C	&	21.80${\pm}$0.08	&	23.71$^{+0.14}_{-0.16}$	&	\hst\ /WFC3	&	F160W$+$F125W	&	\citet{2022MNRAS.tmp..864C}	\\
CXOU\,J1714(05.7--381031)	&	17.45$\pm$0.01	&	19.01$\pm$0.01	&	\hst\ /WFC3	&	F160W$+$F125W	&	\citet{2022MNRAS.tmp..864C}	\\
SGR\,1745(--2900)	&	$>$15.97	&	$>$18.98	&	\hst\ /WFC3	&	F160W$+$F125W	&	\citet{2022MNRAS.tmp..864C}	\\
SGR\,1806(--20)	&	21.75$\pm$0.75	&	 $>$21.1	&	Keck/NIRC2, VLT/NAOS	&	$K_{p}$,$J$	&	\protect{\citet{2012ApJ...761...76T}, \citet{2005A&A...438L...1I}}	\\
XTE\,J1810(--197)$\dagger$	&	21.67$\pm$0.12	&	22.92$\pm$0.22	&	Gemini/NIRI	&	$H$,$J$	&	\protect{\citet{2008A&A...482..607T}}	\\
Swift\,J1822(.3--1606)	&	19.76$\pm$0.01	&	21.04$\pm0.02$	&	\hst\ /WFC3	&	F160W$+$F125W	&	\citet{2022MNRAS.tmp..864C}	\\
Swift\,J1834(.9--0846)	&	21.74$\pm0.02$	&	23.21$\pm0.03$	&	\hst\ /WFC3	&	F160W$+$F125W	&	\citet{2022MNRAS.tmp..864C}	\\
1E\,1841(--045) 	&	20.80${\pm}$0.4	&	 $>$22.10	&	Magellan /Panic	&	$H$,$J$	&	\protect{\citet{2005ApJ...632..563D}}	\\
3XMM\,J1852(46.6$+$003317)$\star$	&	18.85$\pm$0.01	&	21.97$\pm$0.02	&	\hst\ /WFC3	&	F160W$+$F125W	&	\citet{2022MNRAS.tmp..864C}	\\
SGR\,1900($+$14)$\dagger$	&	21.17$\pm$0.50	&	 -	&	Keck/NIRC2	&	$H$	&	\protect{\citet{2012ApJ...761...76T}}	\\
SGR\,1935($+$2154)	&	$\sim$24	&	 -	&	\hst\ /WFC3	&	F140W	&	\protect{\citet{2018ApJ...854..161L}, \citet{2022ApJ...926..121L}}	\\
1E\,2259($+$586)	&	22.52$\pm0.04$	&	23.61${\pm}$0.05	&	\hst\ /WFC3	&	F160W$+$F125W	&	\citet{2022MNRAS.tmp..864C}	\\
SGR\,0755(--2933)$\star\star$	&	9.52$\pm$0.01	&	9.69$\pm$0.01	&	2MASS	&	$H$,$J$	&	\protect{\citet{2021A&A...647A.165D}}	\\
AX\,J1845(.0--0258)	&	$>$21	&	 -	&	NTT	&	$H$	&	\protect{\citet{2004IAUS..218..247I}}	\\
SGR\,2013($+$34)	&	$>$18.5	&	$>$19.3	&	PAIRITEL	&	$H$,$J$	&	\protect{\citet{2005GCN..4042....1B}}	\\
\hline %
\hline %
\end{tabular}
\newline
$\star$ - Reddest object in large error circle, unlikely to be the counterpart \citep{2022MNRAS.tmp..864C}, $\star\star$ - a possibly unrelated HMXB \citep{2021A&A...647A.165D}\\
$\dagger$ - has \hst\ imaging in \citet{2022MNRAS.tmp..864C}, not used due to non-detections or single-band imaging
\end{table*}

\section{Extinctions and distances}\label{sec:corr}
Since the photometry in Table \ref{tab:data} is reported in terms of dust attenuated apparent magnitudes, we must now correct the observed photometry for the distance and dust extinction along the magnetar sight-lines, in order to compare with the un-attenuated, absolute magnitude predictions of BPASS.
 
We adopt the primary distance estimates listed in the McGill catalogue \citep[][and reference therein]{2014ApJS..212....6O}$^{2}$. These are derived using a variety of methods, including neutral hydrogen column densities/red clump stars \citep[e.g.,][]{2006ApJ...650.1070D}, dispersion measures \citep[e.g.,][]{2010ApJ...721L..33L}, and associations with objects that have a well-measured distance \citep[such as supernova remnants, molecular gas clouds, or young clusters][]{1999ApJ...526L..29C,2008MNRAS.386L..23B}. In a few cases the distance is derived  by assuming that the magnetar is in a spiral arm along that line of sight. In this instance, and any others where the distance uncertainty is not quantified, we assume a 15 per cent error (the mean uncertainty of those that are quantified, and large enough to account for spiral arm width). For SGR\,1935, we adopt the estimate of \cite{2021MNRAS.tmp..771B}, which is derived from the dispersion measure and is constrained to $1.5<d<6.5$\,kpc. All distance estimates adopted and their sources are provided in Table \ref{tab:samples}.

To infer the reddening due to dust, $E(B-V)$, we try two methods in the following order of preference,
\begin{enumerate}
    \item If the magnetar lies above declination --30$\degr$, we check the extinction estimate from the Bayestar 3D dust map of \citet[][G19]{2019ApJ...887...93G}, using the {\sc dustmaps} python package \citep{2018JOSS....3..695M}. These maps do not have 100 per cent coverage of the Galaxy, and are only reliable out to a certain distance along each sight-line because stars (in the input surveys \gaia , Pan-STARRS and 2MASS) need to be detected in order to infer an extinction. To determine whether the extinction estimates are reliable, we compare the total cumulative extinction along each sight-line to the 2D extinction estimate from the \citet[][SF11]{2011ApJ...737..103S} dust maps. Where these totals are comparable, we adopt the 3D map estimate at the distance of the magnetar. If there is a large discrepancy (greater than a 50 per cent difference in $A_{\rm V}$) we move on to the next method. An extreme example is SGR\,1745 towards the Galactic centre, where we have $A_{\rm V}\sim3$ from Bayestar and $A_{\rm V}\sim300$ from the 2D map.
    \item The neutral hydrogen column density $N_{\rm H}$ is known to be proportional to visual extinction $A_{\rm V}$. \citet[][PS95]{1995A&A...293..889P} find $A_{\rm V} = N_{\rm H}$/$[(1.79\pm0.03)\times10^{21}\rm{cm}^{2}]$. Many magnetars have an X-ray derived $N_{\rm H}$ estimate, but the uncertainties on this number are typically large, and the $N_{\rm H}$-$A_{\rm V}$ relation itself has some scatter. This method therefore more accurately probes the direct sight-line to the magnetar, but is typically less precise.
\end{enumerate}

For the first method, the uncertainty on the extinction is correlated with the uncertainty on the distance. For example, if a 3D estimate is used, the upper bound on the extinction is the dust map value at $d+{\delta}d$. The $N_{\rm H}$ inferred $A_{\rm V}$ values depend on X-ray measurements and we treat them independently from the distance estimate. 

\begin{figure}
	\includegraphics[width=0.99\columnwidth]{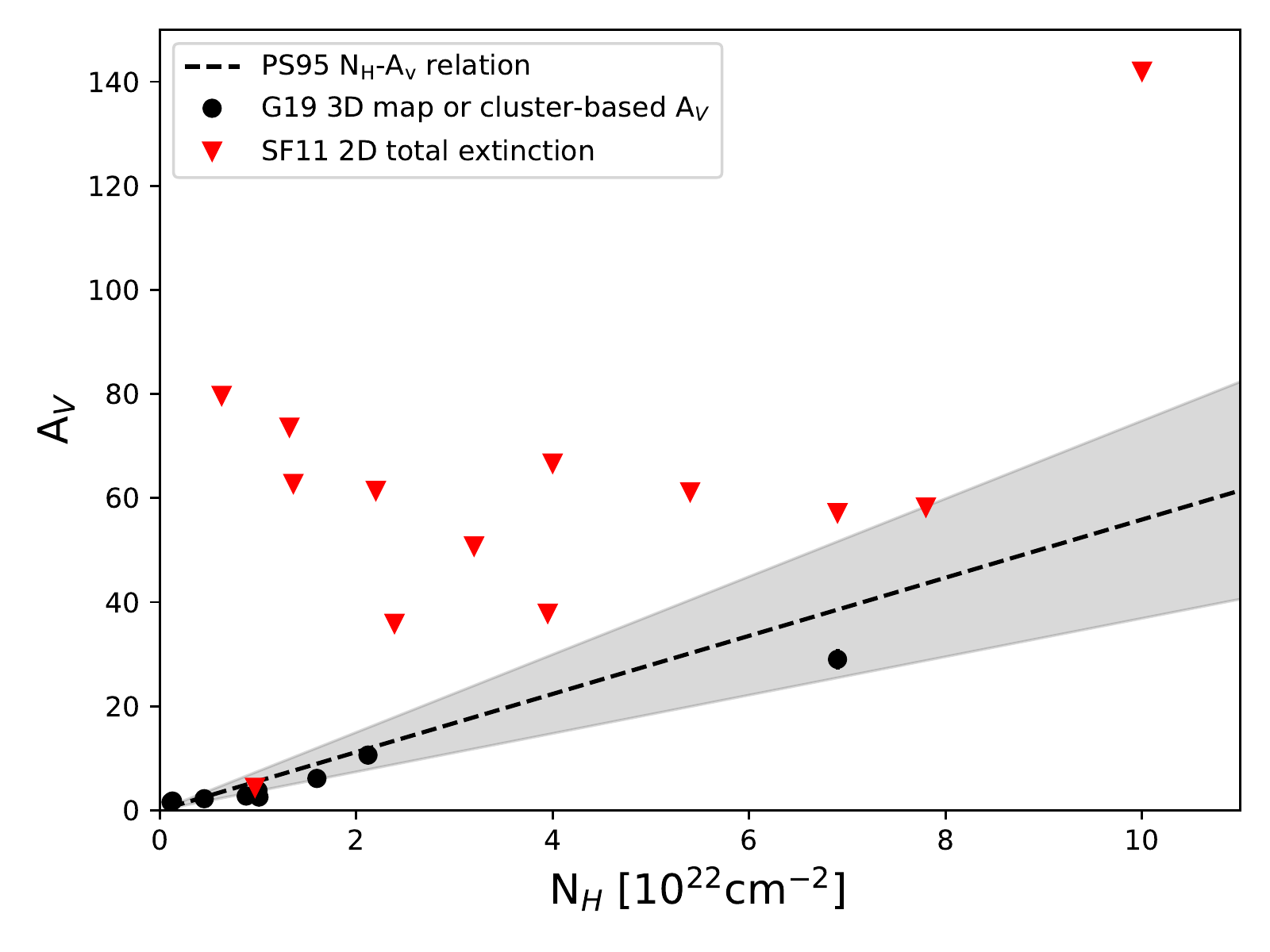}
    \caption{Extinction as a function of N$_{\rm H}$ derived from (i) magnetar distance estimates and the 3D maps of \citet[][G19]{2019ApJ...887...93G}, or a cluster association-based extinction estimate if available, (ii) the N$_{\rm H}$-$A_{\rm V}$ relation of \citet[][PS95]{1995A&A...293..889P} and (iii) the total integrated line of sight extinction out of the Galaxy \citet[][SF11]{2011ApJ...737..103S}. We adopt 3D dust map estimates if they are reliable, if not and no cluster-based estimate is available, then the $N_{\rm H}-A_{\rm V}$ relation is used. The shaded region around the A$_{\rm V}$-N$_{\rm H}$ relation corresponds to the typical $\sim$0.3 relative difference between 3D dust map and N$_{\rm H}$ derived estimates.} 
    \label{fig:avnh}
\end{figure}

In Figure \ref{fig:avnh} we plot $N_{\rm H}$ versus $A_{\rm V}$ for each magnetar with a $N_{\rm H}$ measurement. If a 3D dust map or cluster-based extinction is used, that value is plotted. For example, we adopt a value based on observations of the Wd1 cluster for CXOU\,J1647 \citep{2014A&A...565A..90C}. If an $A_{\rm V}$-$N_{\rm H}$ value is adopted, the 2D value for that sight-line is indicated. The \citet{1995A&A...293..889P} relation is also shown. The quoted scatter on the $A_{\rm V}$-$N_{\rm H}$ relation is a significant underestimate, if the differences between the 3D dust map values and $N_{\rm H}$ derived values are representative. The difference might partly arise due to gas and dust local to the magnetars, not captured by the dust map spatial resolution (variable, but between 3-14\,arcmin for Bayestar), explaining the tendency for the 3D dust map values to be lower than the $N_{\rm H}$ estimates. The standard deviation of the relative differences between the 3D dust map and N$_{\rm H}$ derived extinctions, i.e. $\Delta A_{\rm V}$/$A_{\rm V}$, is 0.34. This is indicated in Figure \ref{fig:avnh} by the grey shaded region on the $A_{\rm V}$-$N_{\rm H}$ relation. We plot this region equally either side of the relation, although there is a clear bias towards $A_{\rm V}$-$N_{\rm H}$ values exceeding dust map values.

The total 2D extinctions greatly exceed the N$_{\rm H}$ estimates as expected, bar one case at low $A_{\rm V}$, putting an upper limit on the degree to which local, unresolved dust/gas enhancement is contributing to the higher N$_{\rm H}$ derived estimates. Since the N$_{\rm H}$ and 3D map methods agree to within $\sim$30 per cent, this also implies that the distance estimates are reasonable, since the extinction is a function of distance for the 3D map estimates. In each case, we choose the most reliable method as described above, where the 3D map values are correlated with distance and both the quoted scatter and N$_{\rm H}$ uncertainties are included if \citet{1995A&A...293..889P} is used. We note that in both cases, the extinction uncertainties may be underestimated, but that they will be no more than $\sim$30 per cent in any case.

Finally, we convert visual $A_{\rm V}$ extinctions into $H$,$J$, F160W, F140W, F125W and F110W extinctions using the effective wavelengths of these filters and the python {\sc extinction} module \citep{barbary_kyle_2016_804967}, assuming R$_{\rm V}=3.1$ and a Fitzpatrick extinction curve \citep{1999PASP..111...63F}. The extinction values using each method, and the adopted values, are listed in Table \ref{tab:samples}.

\section{Comparison with population synthesis predictions}\label{sec:compare}
We now compare the observed colours and magnitudes of the candidate bound companions to the predictions from Section \ref{sec:bpass}.

\subsection{H-band magnitude comparisons}
\begin{figure*}
	\includegraphics[width=0.99\textwidth]{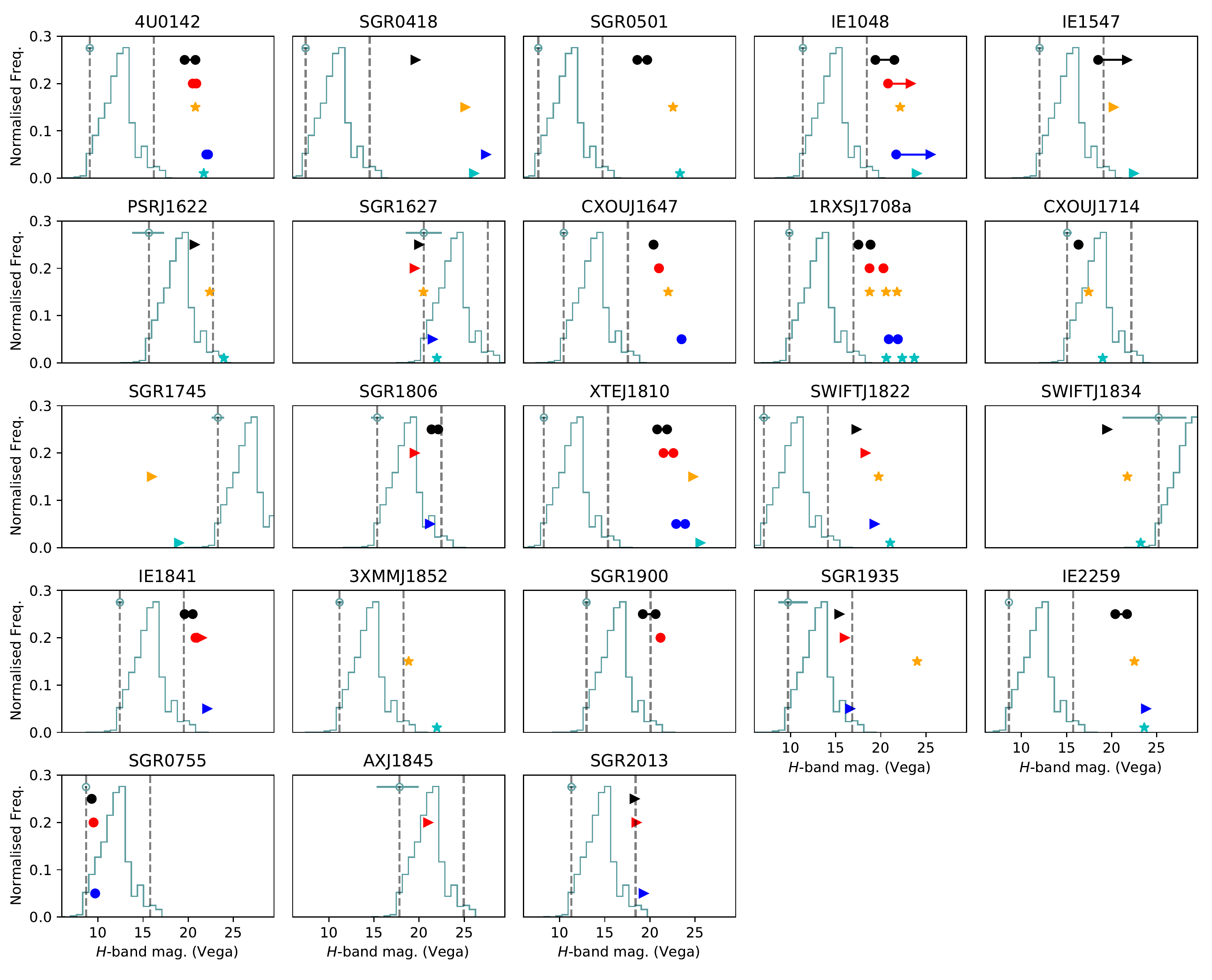}
    \caption{Photometry for the 23 magnetars with a distance estimate, compared to BPASS predictions for the apparent magnitudes of bound companions (histogram), assuming the distance and extinction estimates in Table \ref{tab:samples}. The 2.5 and 97.5 percentiles (enclosing 95 per cent of the distribution) are indicated by vertical lines, the lower percentile has an errorbar indicating the uncertainty on the distribution due to distance and $A_{\rm V}$ uncertainties. Previously reported apparent magnitudes (uncorrected) are indicated by the black points ($K$-band, arbitrarily placed at $y=0.25$), red ($H$-band, 0.2) and blue ($J$-band, 0.05). If the source is variable and a range of values have been reported, the brightest and faintest magnitudes previously reported are joined by a line. If two separate sources have previously been measured as candidates, they are plotted separately. Triangles represent upper limits. The \hst\ F160W (F125W) photometry listed in Table \ref{tab:samples} is plotted in orange (cyan) at $y=0.15$ ($y=0.01$).}
    \label{fig:bound_H}
\end{figure*}

Using the distance estimates in Table \ref{tab:samples}, the absolute magnitudes of NIR counterpart candidates can be calculated. As a first step in comparing the predictions of Section \ref{sec:bpass} to observations, we instead correct the $H$-band absolute magnitude distribution for bound companions (Figure \ref{fig:massH}) for the distance and extinction of each magnetar. These dust attenuated apparent magnitudes are compared to observations in Figure \ref{fig:bound_H}. Dashed vertical lines enclose 95 per cent of the $H$-band magnitude probability. Uncertainties on the correction to the BPASS outputs are indicated by error-bars on the lower (brighter) percentile. In calculating these uncertainties (provided in Table \ref{tab:samples}), we assume for dust-map derived extinctions that the distance and extinction uncertainties are correlated, i.e. the upper distance error is used with the upper extinction error to derive the maximum correction within the uncertainties, and vice versa. In every other case we assume that the uncertainties are uncorrelated, taking the average of the upper and lower distance and extinction uncertainties if they are asymmetric. The observed magnitudes are either from \hst\ measurements (F140W is drawn on this figure as F160W, F110W as F125W) or other observations from the literature (if not \hst\, they are assigned to $H$,$J$ or $K$, whichever is the closest match) as indicated in Table \ref{tab:data}. Although SGR\,1806 does not have a $H$-band detection, if we assume it has a similar intrinsic $H$--$K=1$ colour to other counterparts, combined with 2 magnitudes more extinction in $H$ than $K$ ($A_{\rm V}=$29), this places it a few magnitudes outside the lower (in terms of luminosity) percentile.

\begin{table*}
\caption{Details for each of the 23 magnetars with a distance estimate and imaging, including distance estimates, and extinctions calculated using the two methods outlined in the text. The majority of this information is as provided in the McGill magnetar catalogue and references therein \citet{2014ApJS..212....6O}$^{2}$, with some additions and updates. The extinction used for each magnetar is in bold, and listed in the final column with the uncertainty. For dust-map derived extinctions, the upper extinction uncertainty corresponds to the upper distance uncertainty, and vice versa.
}
\label{tab:samples}
\begin{tabular}{lllllll}
\hline
Magnetar	&	Dist.	&	Distance reference	&	Total $A_{\rm V}$	&	$A_{\rm V}$ at $d$	&	$A_{\rm V}$-$N_{\rm H}$	&	$A_{\rm V}$ used	\\
	&	[kpc]	&		&	SF11 (tot. G19)	&	G19	&	PS95	&	with uncertainty	\\
\hline													
4U\,0142	&	3.6$\pm$0.4	&	\protect{\cite{2006ApJ...650.1070D}}	&	4.90 (4.93)	&	{\bf 3.79}	&	5.59	&	3.79$_{-0.25}^{+0.40}$	\\
SGR\,0418	&	2$\pm$0.3${\dagger}$	&	\protect{\cite{2010ApJ...711L...1V}}	&	2.25 (2.17)	&	{\bf 1.55}	&	0.64	&	1.55$\pm$0.01	\\
SGR\,0501	&	2$\pm$0.3${\dagger}$	&	\protect{\cite{2011ApJ...739...87L}}	&	3.45 (3.62)	&	{\bf 2.73}	&	4.92	&	3.79$_{-0.39}^{+0.15}$	\\
IE\,1048	&	9.0$\pm$1.7	&	\protect{\cite{2006ApJ...650.1070D}}	&	3.75	&	 -	&	\bf{5.42}	&	5.42$\pm$0.11	\\
1E\,1547	&	4.5$\pm$0.5	&	\protect{\cite{2010ApJ...710..227T}}	&	43.42	&	 -	&	\bf{17.88}	&	17.88$_{-0.11}^{+1.15}$	\\
PSR\,J1622	&	9$\pm$1.35${\dagger}$	&	\protect{\cite{2010ApJ...721L..33L}}	&	53.22	&	 -	&	\bf{30.17}	&	30.17$_{-7.84}^{+8.94}$	\\
SGR\,1627	&	11$\pm$0.3	&	\protect{\cite{1999ApJ...526L..29C}}	&	125.4	&	 -	&	\bf{55.87}	&	55.87$_{-11.22}^{+11.21}$	\\
CXOU\,J1647	&	3.9$\pm$0.7	&	\protect{\cite{2007A&A...468..993K}}	&	30.61	&	 \bf{10.66} $[1]$	&	13.35	&	10.66$\pm$0.40	\\
1RXS\,J1708	&	3.8$\pm$0.5	&	\protect{\cite{2006ApJ...650.1070D}}	&	53.84	&	 -	&	\bf{7.60}	&	7.60$_{-0.26}^{+0.25}$	\\
CXOU\,J1714	&	13.2$\pm$1.98${\dagger}$	&	\protect{\cite{2012MNRAS.421.2593T}}	&	31.36	&	 -	&	\bf{22.07}	&	22.07$_{-0.87}^{+0.91}$	\\
SGR\,J1745	&	8.3$\pm$0.3	&	\protect{\cite{2014ApJ...780L...2B}}	&	301.66 (3.34)	&	unreliable	&	 \bf{75.42}	&	75.42$_{-3.53}^{+3.58}$	\\
SGR\,1806	&	8.7$^{+1.8}_{-1.5}$	&	\protect{\cite{2008MNRAS.386L..23B}}	&	45.5 (6.85)	&	\bf{29 [2]}	&	38.55	&	29$\pm$2	\\
XTE\,J1810	&	2.5$^{+0.4}_{-0.3}$	&	\protect{\cite{2020MNRAS.498.3736D}}	&	60.05 (5.43)	&	unreliable	&	\bf{3.52}	&	3.52$\pm$0.29	\\
Swift\,J1822	&	1.6$\pm$0.3	&	\protect{\cite{2012ApJ...761...66S}}	&	12.75 (9.73)	&	\bf{2.21}	&	2.53	&	2.21$_{-1.32}^{+1.18}$	\\
Swift\,J1834	&	4.2$\pm$0.3	&	 \protect{\cite{2008AJ....135..167L}}	&	96.78 (4.55)	&	unreliable	&	 {\bf 98.27}	&	98.27$_{-20.13}^{+19.11}$	\\
IE\,1841	&	8.5$^{+1.3}_{-1.0}$	&	\protect{\cite{2008ApJ...677..292T}}	&	52.86 (5.61)	&	unreliable	&	\bf{12.29}	&	12.29$_{-0.59}^{+0.60}$	\\
3XMM\,J1852${\ddagger}$	&	7.1$\pm$1.07${\dagger}$	&	\protect{\cite{2014ApJ...781L..16Z}}	&	65.17 (9.89)	&	unreliable	&	 {\bf 7.63}	&	7.63$_{-0.56}^{+0.05}$	\\
SGR\,1900 	&	12.5$\pm$1.7	&	\protect{\cite{2009ApJ...707..844D}}	&	11.05 (10.94)	&	\bf{10.60}	&	11.84	&	10.60$_{-0.03}^{+0.07}$	\\
SGR\,1935	&	4.0$\pm2.5$	&	\protect{\cite{2021MNRAS.tmp..771B}}	&	13.13 (8.58)	&	\bf{6.11}	&	8.94	&	6.11$_{-0.31}^{+0.28}$	\\
IE\,2259	&	3.2$\pm$0.2	&	\protect{\cite{2012ApJ...746L...4K}}	&	3.82 (3.22)	&	\bf{2.54}	&	5.64	&	2.54$_{-0.03}^{+0.04}$	\\
SGR\,0755	&	3.5$\pm$0.2	&	 \protect{\cite{2021A&A...647A.165D}}	&	2.6 (2.39)	&	\bf{1.77}	&	0.73	&	1.77$_{-0.03}^{+0.09}$	\\
AX\,J1845	&	8.5$\pm$1.28	&	\protect{\cite{1998ApJ...503..843T}}	&	50.35 (5.58)	&	unreliable	&	\bf{43.58}	&	43.58$_{-10.09}^{+12.87}$	\\
SGR\,2013	&	8.5$\pm$1.32	&	\protect{\cite{2011AdSpR..47.1346S}}	&	5.64 (6.20)	&	\bf{5.36}	&	6.13	&	5.36$_{-1.24}^{+0.56}$	\\
\hline %
\hline %
\end{tabular}
\newline
${\dagger}$ - distance uncertainty unquantified, 15 per cent assumed, ${\ddagger}$ - Reddest source in error circle \\
$[1,2]$ - extinctions from cluster associations \citep{2014A&A...565A..90C,2004ApJ...616..506E}.\\
\end{table*}

\begin{figure*}
	\includegraphics[width=0.49\textwidth]{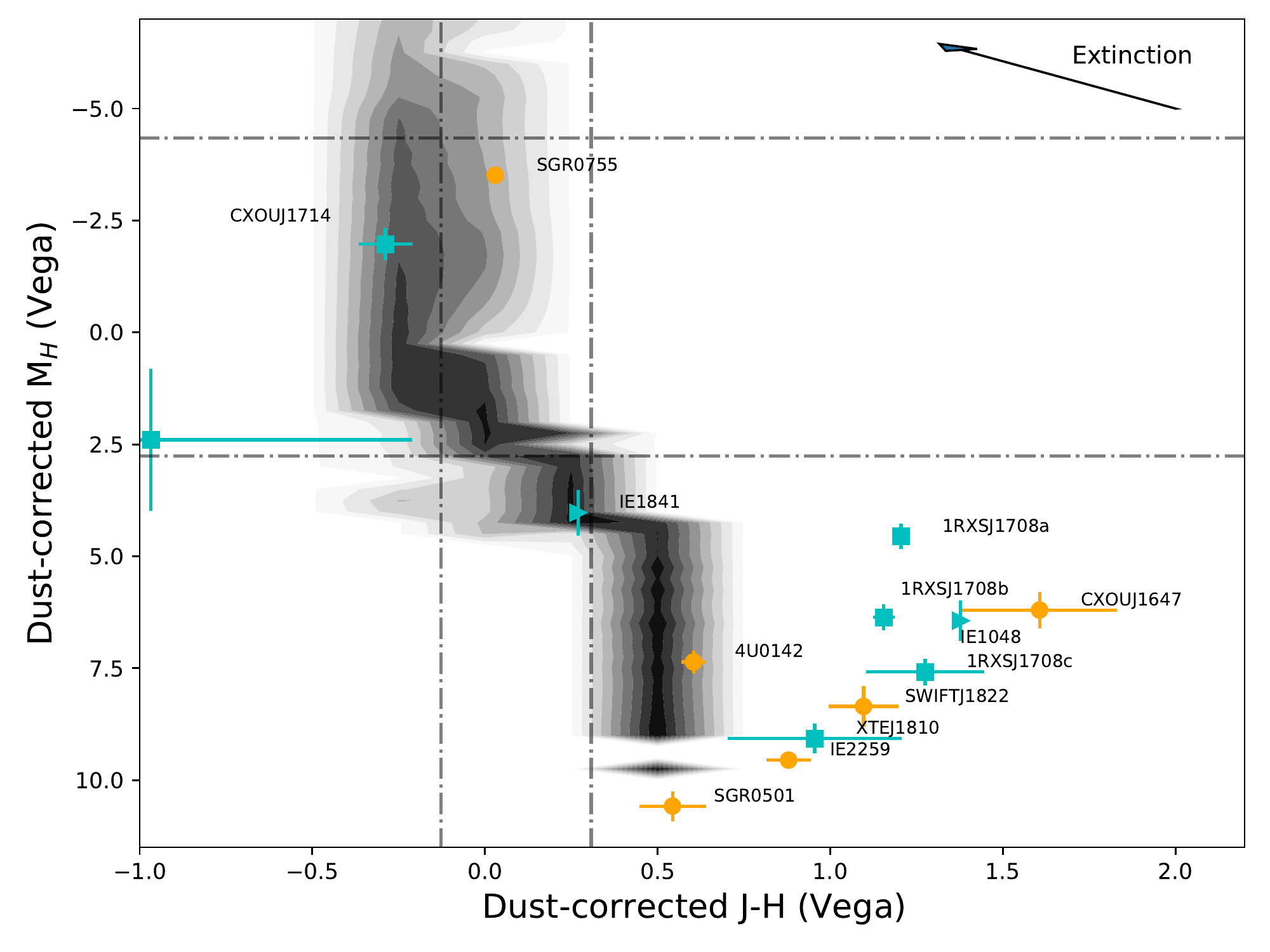}
	\includegraphics[width=0.49\textwidth]{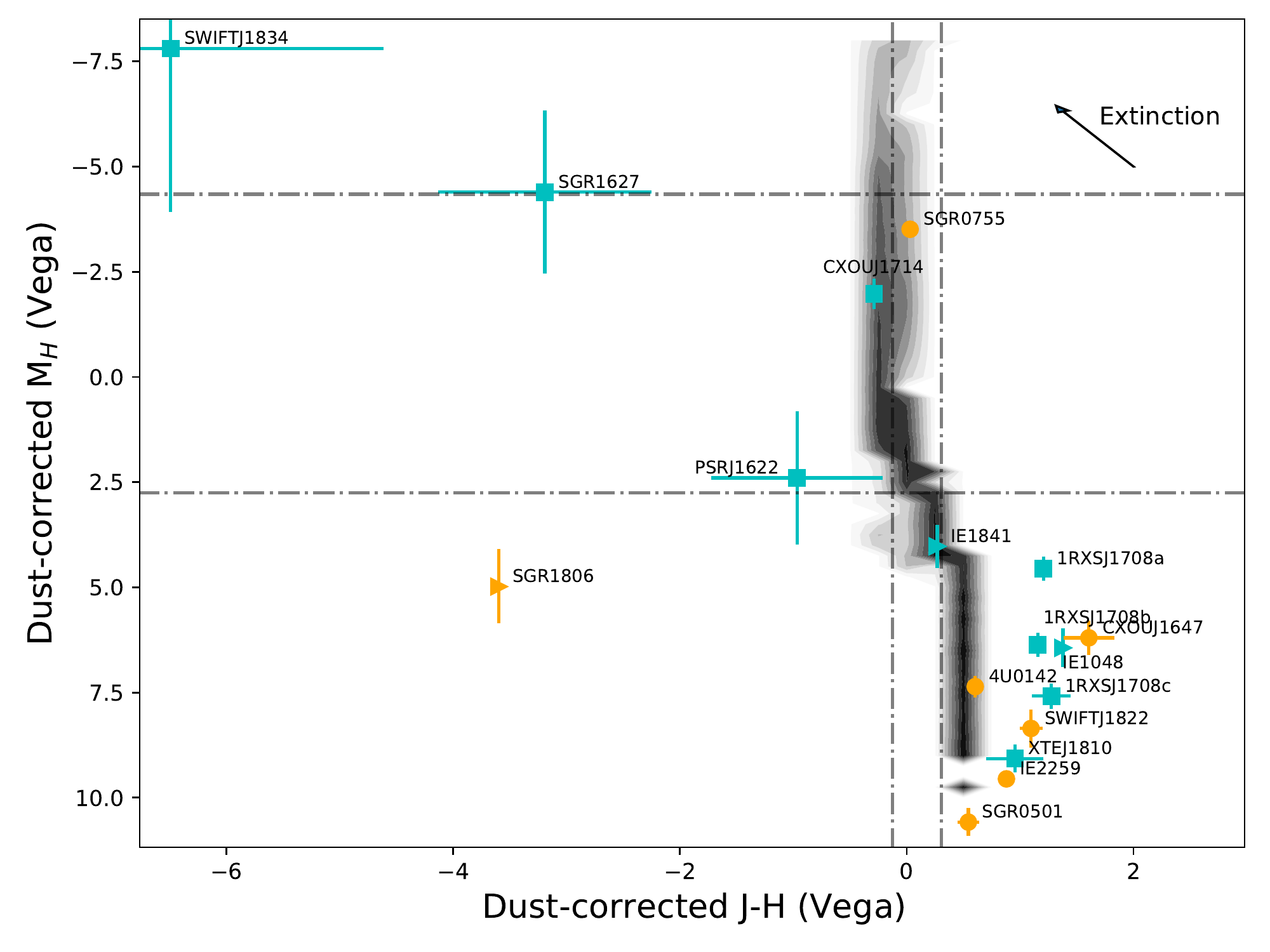}
    \caption{A NIR colour magnitude diagram for a Solar metallicity stellar population at 10\,Myr. The right hand panel zooms out to show SGR\,1806, PSR\,J1622, SGR\,1627 and Swift\,J1834 (whose extinctions are large and uncertain). Extinctions derived from 3D dust map/cluster extinctions, or the $A_{\rm V}$-N$_{\rm H}$ relation, are labelled with orange circles or cyan squares respectively. Immediately post primary supernova, 95 per cent of all bound companion $H$-band magnitudes lie in the region bounded by the horizontal dot-dashed lines, and 95 per cent of their $J$--$H$ colours lie between the vertical dashed lines. The reddening vector is indicated in the upper right - for a fixed apparent magnitude and observed colour, the inferred absolute magnitude and intrinsic colour move in this direction with increasing extinction. Objects that lie away from the shaded regions either have intrinsically non-stellar colours, or the incorrect distance and extinction applied (i.e., they are chance alignments). The underlying density contours (an order of magnitude in number density per level) were plotted with the python package for BPASS, HOKI \citep{2020JOSS....5.1987S}.} 
    \label{fig:cmd}
\end{figure*}

\subsection{Colour-magnitude comparisons}
While informative, single-band magnitudes are only part of the information available. To add further constraints, we now examine the colours of the counterpart and candidates. For those sources which have observations in $H$ and $J$ and a detection in at least one, we apply the distance and extinction corrections to the {\it magnetar} photometry and place this on a colour-absolute magnitude diagram in Figure \ref{fig:cmd}. Here we show BPASS predictions for a Solar metallicity stellar population at 10$^{7}$\,yrs. Ages of a fews tens of Myr correspond to the expected lifetimes of supernova progenitors, hence our choice of 10$^{7}$\,yr, but comparisons with populations at 10$^{6}$\,yr and 10$^{8}$\,yr are also provided in Appendix \ref{apx:cmdages}.

The corrected magnetar measurements are coloured by the method used to derive the extinction. Horizontal dashed lines mark the upper and lower luminosity percentiles enclosing 95 per cent of the distribution, previously adjusted in each case in Figure \ref{fig:bound_H}. Vertical dashed lines represent the same for the predicted $J$--$H$ colour distribution of bound companions. The box enclosed by these 4 lines is therefore where companion stars are predicted. However, we plot the whole stellar population at the given age underneath for two reasons. Firstly, this is in order to be agnostic about the properties of any bound companions, which represent a subset of the wider stellar population. For example, as companions to  massive stars, they will be biased to higher masses since very high mass ratio systems are disfavoured \citep{2017ApJS..230...15M}. However, by plotting the whole population, we are open to the possibility of companions outside of the predicted range, either because the predictions are inaccurate, or because magnetar progenitor systems are atypical. Secondly, if a source does not fall on the shaded regions (where each shading level represents an order of magnitude difference in the number density of sources), this is indicative of either,
\begin{itemize}
    \item Genuinely non-stellar colours, assuming the adopted $d$ and $A_{\rm V}$ are at least approximately correct, or,
    \item A chance alignment, where the source in the localisation error circle is unrelated to the magnetar, and hence the wrong distance and extinction is being applied,
\end{itemize}
where `non-stellar' is defined as colour-magnitude properties which are inconsistent with stars undergoing nuclear fusion (any phase from the main sequence through to core-collapse or white dwarf formation), or white dwarfs. Examples of non-stellar emission in this context include magnetospheric or debris disc emission. We now discuss the placement of the counterparts in this parameter space after distance and extinction correction, and the implications in each case.

There are three cases where the wrong corrections have likely been applied for the sources in the error circle. These are SGR\,1627, PSR\,J1622 and Swift\,J1834, which have very large uncertainties due to large and uncertain N$_{\rm H}$ measurements. None the less, they are unphysically blue (the right panel of Fig. \ref{fig:cmd} is zoomed out to show this). Given the high extinction on these sight-lines, it is likely that these are foreground chance alignments, and therefore the distance and extinction corrections are much too high. This is also seen by the direction of the extinction vector, which traces back to the main sequence - the region of highest probability density - for these sources.

At first glance, the three sources in (or just outside, we have added source B of \citet{2003ApJ...589L..93I} and \citet{2006ApJ...648..534D} to compare) the 1RXS\,J1708 error circle also look like chance alignments, since they cannot all be counterparts. However, \cite{2003ApJ...589L..93I} claim that their source A (the brightest of the three) is non-stellar in $J$--$H$, $J$--$K$ colour space. \citet{2006ApJ...648..534D} showed that source B can only be shifted into the cloud of stellar sources if an $A_{\rm V}$ of ${\sim}$20 is assumed, inconsistent with the column density extinction estimate (and consistent with the main sequence offset shown in Figure \ref{fig:cmd}). It is therefore unclear which is associated with 1RXS\,J1708, but it is likely that two are anomalously red because they are background chance alignments with an insufficient distance/extinction applied, whereas the other is the magnetar counterpart, with a non-stellar SED. AX\,J1818.0-1607 is a similar case, for which this was shown to be a plausible scenario by \citet{2022MNRAS.tmp..864C}. In any case, all of the 1RXS\,J1708 candidates are outside the bound companion magnitude range.

There are another nine sources below the population synthesis bound companion limit. 4U\,0142, 1E\,2259 and SGR\,0501 are unambiguously the counterpart based on precise localisations and a lack of other sources in the immediate vicinity. 1E\,2259 stands out as being unusually red, whereas 4U\,0142 does not, despite previous claims of a dusty debris disc. We attribute this to the $J$ and $H$ bands used, which are not as sensitive in this case to the upturn which becomes more significant from the $K$-band into the mid-infrared \citep[e.g.][]{2016MNRAS.458L.114M}. The source associated with Swift\,J1822 is also in this faint, red area of parameter space. 

1E\,1841, 1E\,1048 and SGR\,1806 all have $J$-band non-detections and hence their colours are lower limits. For 1E\,1048 this is a constraining limit which places it in the same faint, red region of parameter space as the other counterparts. 1E\,1841 has $H$ and $K$ imaging presented by \cite{2008A&A...482..607T} and the likely counterpart is identified as their source 9, due to 4$\sigma$ $K$-band variability. We therefore adopt the $H$ and $J$-band measurements of the same object in \citet{2005ApJ...632..563D}, labelled there as source B. The shallow $J$-band limit means that the $J$--$H$ colour is relatively poorly constrained. The source placement is outside of the predicted companion magnitude range and marginally consistent with the predicted colours. For SGR\,1806, which has a $K$-band detection \citep{2012ApJ...761...76T} and limits in $H$ and $J$ \citep{2005A&A...438L...1I}, we first assume a $H$--$K$ colour of one. This is typical of the intrinsic colours of known counterparts (see e.g. the least extincted examples in Figure \ref{fig:bound_H}) and is consistent with the $H>19.5$ limit. Correcting for the distance and extinction then places it at M$_{H}\sim$5. Given the $J>21.1$ limit, the high extinction ($A_{\rm V}=29$) yields an unconstraining $J$--$H>-3.6$ colour limit.

Although there is \hst\ imaging for CXOU\,1647, this is only in one band (F140W, \citet{2022MNRAS.tmp..864C}), so we adopt the $J=23.5\pm0.2$ and  $H=21.0\pm0.1$ measurements of \citep{2018MNRAS.473.3180T} instead. The last of the eight sources with colour information to lie below the population synthesis prediction for bound companion $H$-band magnitudes is XTE\,J1810. The counterpart is not detected in 2018 \hst\ F160W and F125W imaging \citep{2022MNRAS.tmp..864C}, so we use measurements from \citet{2008A&A...482..607T} instead. Because the source is variable, we chose just the first epoch (MJD 52900) of observations presented ($H=21.67\pm0.12$, $J=22.92\pm0.22$).

Finally, there are two sources which are notably different from the cloud of objects in the lower right of Figure \ref{fig:bound_H}. Not only do they lie in the magnitude range predicted for companions, but they also have stellar colours. These are SGR\,0755 and CXOU\,J1714. The star associated with SGR\,0755, as previously discussed, is too bright to measure from the \hst\ images, so 2MASS \citep{2006AJ....131.1163S} magnitudes are adopted. This system, 2SXPS\,J075542.5--293353, is an HMXB \citep{2021A&A...647A.165D}. However, it remains an open question whether the accretor could be a magnetar \citep{2022arXiv220107507P}, or whether the magnetar burst came from a different source in the BAT error circle. The counterpart candidate for CXOU\,J1714 is unambiguous, well localised, and has stellar colours, albeit slightly bluer than predicted for neutron star companions. The source also stands apart from previously established counterparts in terms of its X-ray to NIR power law index \citep{2022MNRAS.tmp..864C}. Unless the distance and extinction are drastically overestimated for this magnetar, the source is clearly distinct from previously confirmed non-stellar counterparts.

\section{Summary of results}\label{sec:summary}
Out of the 23 sources considered in this sample, 3 are unconstrained in terms of their $H$-band magnitude - AX\,J1845 only has a limit that does not preclude a bound companion, SGR\,1745 also has an unconstraining limit, and we deem 3XMM\,J1852 as unconstrained given the poor localisation. The overall results, including colour information where available, are summarised in Table \ref{tab:results}. Here we list whether the observations for each source can rule out or allow for the possibility of a bound companion. Of the 20 that have a magnitude constraint, 2 are plausibly bound companions based on their absolute magnitude and colour. These are SGR\,0755 and CXOU\,J1714. CXOU\,J1714 is slightly outside the colour predictions but is otherwise well within the magnitude range expected. SGR\,0755 satisfies both criteria. IE\,1841 is not far ($\sim 2\sigma$) outside the lower magnitude bound, but the colour limit also precludes most of the $J$--$H$ distribution, so we deem it inconsistent with predictions.

Of the two candidates, which appear distinctly bright and blue in this parameter space, SGR\,0755 is an X-ray binary, and it is possible that the magnetar is actually elsewhere in the BAT error circle \citep{2021A&A...647A.165D}. CXOU\,J1714 is perhaps the most promising bound companion candidate. We can therefore constrain the fraction of magnetars with a bound companion to 0.05--0.10 (1--2 out of 20). This range is quoted at 95 per cent confidence since both the localisation uncertainties and bound companion magnitude predictions are at 95 per cent. This bound companion fraction is consistent with population synthesis predictions for a regular core-collapse origin, as discussed in Section \ref{sec:bpass}. If there is a contribution amongst this population from non-core-collapse channels, it cannot be very large before the bound companion fraction among the remainder disagrees with predictions. For instance, if 5/20 do not have a core-collapse origin, the bound fraction amongst the rest would be 7--13 per cent, which would imply a bias towards magnetar production in primary stars, assuming that $f_{\rm bound}=0.05$ is accurate for the wider neutron star population.

\begin{table}
\caption{A summary of the results. Whether each source is plausibly a bound companion based on the $H$-band or location on the colour-magnitude diagram (CMD), is indicated. The tolerance for being consistent in Figure \ref{fig:bound_H} is extended beyond the plotted percentiles according to the indicated uncertainty on the distance and extinction corrections. 3/23 are unconstrained. Of the remaining 20, two have apparently stellar properties broadly consistent with expectations for bound companions (CXOU\,J1714 and SGR\,0755, for which the bright star is a confirmed HMXB, but may but a chance alignment with the magnetar in question).} 
\label{tab:results}
\begin{tabular}{llll}
\hline %
Magnetar	&	Plausible	&	Plausible	&	Classification	\\
	&	from H mag?	&	from CMD?	&		\\
\hline							
4U\,0142	&	N	&	N	&	Non-stellar	\\
SGR\,0418	&	N	&	 -	&	Non-stellar	\\
SGR\,0501	&	N	&	N	&	Non-stellar	\\
1E\,1547	&	N	&	 -	&	Non-stellar	\\
IE\,1048	&	N	&	 N	&	Non-stellar	\\
PSR\,J1622	&	Y	&	N	&	Non-stellar${\star\star}$ 	\\
SGR\,1627	&	Y	&	N	&	Non-stellar${\star\star}$ 	\\
CXOU\,J1647	&	N	&	N	&	Non-stellar	\\
1RXS\,J1708	&	N	&	N	&	Non-stellar	\\
CXOU\,J1714	&	Y	&	Y	&	Possibly stellar	\\
SGR\,J1745	&	Unconstrained	&	 -	&	Unconstrained	\\
SGR\,1806	&	N$^{\star}$	&	 N	&	Non-stellar	\\
XTE\,J1810	&	N	&	 N	&	Non-stellar	\\
Swift\,J1822	&	N	&	N	&	Non-stellar	\\
Swift\,J1834 &	N	&	 N	&	Non-stellar$^{\star\star}$	\\
IE\,1841	&	N	&	 N	&	Non-stellar	\\
3XMM\,J1852	&	Unconstrained	&	 -	&	Unconstrained	\\
SGR\,1900	&	N	&	 -	&	Non-stellar	\\
SGR\,1935	&	N	&	 -	&	Non-stellar	\\
IE\,2259	&	N	&	N	&	Non-stellar	\\
SGR\,0755	&	Y	&	Y	&	Stellar	\\
AX\,J1845	&	Unconstrained	&	 -	&	Unconstrained	\\
SGR\,2013	&	N	&	 -	&	Non-stellar	\\
\hline %
\hline %
\end{tabular}
\newline
${\star}$ - Assuming an intrinsic $H$--$K=1$\newline
${\star\star}$ - extremely blue unattenuated colours, high likelihood of chance alignment
\newline
\end{table}

\section{Discussion}\label{sec:discuss}

\subsection{Bound companion candidates}
\subsubsection{SGR0755}
A system of particular interest here is 2SXPS\,J075542.5--293353, a HMXB which may (or may not) be harbouring SGR\,0755 \citep{2021A&A...647A.165D}. If this is a magnetar accreting from a high mass companion, it would be the first such system identified \citep[some works claim that such systems are highly unlikely to form, since accretion would dampen the magnetic field, or at least that accretion does not power magnetar emission e.g.][]{2019MNRAS.485.3588K,2020A&A...643A.173D}. While it may be a chance coincidence that this HMXB lies in the BAT error circle derived from the SGR\,0755 discovery burst, it could also be that the burst came from this system even if does not contain a magnetar \citep[there are several X-ray pulsars known in HMXBs e.g.][]{2021MNRAS.tmp.3058S}. There are also models which predict magnetar survival (i.e. avoiding rapid magnetic field decay) in certain accreting systems \citep{2018MNRAS.473.3204I,2022MNRAS.510.4645B}. For example, the orbital and spin periods measured for 2SXPS\,J075542.5--293353 agree with the magnetar X-ray binary model predictions of \cite{2021arXiv211010438X}. Unfortunately, we cannot compare the age of the magnetar in this system with an estimate age of the star, because the $P/\dot{P}$ age becomes unreliable in an accreting system. Furthermore, the lack of an independent {\it magnetar} distance estimate precludes a comparison with the \gaia\ distance of 3.5$\pm$0.2\,kpc \citep[EDR3,][]{2016A&A...595A...1G,2021A&A...649A...1G}.

\subsubsection{CXOUJ1714}
The other magnetar with a coincident NIR source consistent with being a massive stellar companion is CXOU\,J1714. It has an inferred absolute magnitude consistent with BPASS predictions for a bound companion, and although bluer than the predicted colours, it is still consistent with being stellar in nature. This source is much fainter than the donor in 2SXPS\,J075542.5--293353, and has not yet been unambiguously identified as stellar in nature. In principle, there might be an orbital period signature in the X-ray lightcurve of the magnetar, if the emission is somehow coupled to interactions with the companion. Figure \ref{fig:massH} shows the BPASS prediction for the orbital periods of bound neutron star-massive companion systems. X-ray observations of CXOU\,J1714 are somewhat sparse, with 11 epochs since its identification as a magnetar in 2010 \citep{2010PASJ...62L..33S}. There are also a handful of preceding epochs, where the target was the surrounding supernova remnant CTB 37B \citep{2019ApJ...882..173G}. The 3.8s period of the magnetar is clear, and there is clear evidence for variability over timescales of $\sim$years within these data \citep{2010PASJ...62L..33S}. However, whether there is day-week timescale periodicity in this long term variation, as predicted in a bound system, remains to be seen. 

We also search for catalogues containing the NIR source associated with CXOU\,J1714. A \gaia\ star lies 2.7\,arcsec away from the centre of the \cxo\ localisation of CXOU\,J1714, and appears in the \gaia\ - neutron star association catalogue of \citet{2021MNRAS.501.1116A}. No proper motion or parallax is available for this source, due to its faint 20.9 $G$-band magnitude. Given the good \cxo\ localisation of the magnetar (ruling the star out as a bound companion), and given the extinction of $A_{\rm V}\sim22$ along the magnetar sight-line, it would have an unphysically luminous absolute $G$-band magnitude of $\sim$--10 if placed at the same distance and extinction as the magnetar (i.e., as a companion).

The only optical/NIR catalogue in which the NIR source at the position of CXOU\,J1714 is detected, is the VVV Infrared Astrometric Catalogue \citep{2018MNRAS.474.1826S}. It has VVV magnitudes of $H=17.3\pm0.1$ and $J=19.1\pm0.3$, agreeing well with our measurements (see Table \ref{tab:samples}), given the bandpass differences. The source has a VVV proper motion measurement of $6.3\pm3.1$\,mas\,yr$^{-1}$. Following \citet{2017A&A...608A..57V} and using the same model parameters as \cite{2022ApJ...926..121L}, we correct for the relative motion due to Galactic rotation and the peculiar velocity of the Sun. The IAU standard of 220\,kms$^{-1}$ is assumed for the Galactic rotation at both the Solar location and the source \citep[reasonable given that all locations considered are outside the central few kpc of the Galaxy,][]{1985ApJ...295..422C}. The resultant transverse velocity in the local standard of rest (LSR) of the source is $349\pm151$\,kms$^{-1}$, where the 1\,$\sigma$ uncertainty is derived from 10$^{5}$ Monte Carlo trials with Gaussian uncertainties on U, V and W (the Solar peculiar motion components), $\mu_{\alpha}$ and $\mu_{\delta}$ (the observed proper motion components) and distance $d$. This is a high value which is more typical of the Galactic pulsar population, but the lower limit is consistent with the upper end of compact remnant-massive star bound systems, specifically high-mass X-ray binaries \citep[][]{2002A&A...395..595M,2019MNRAS.489.3116A,2021MNRAS.508.3345I}. If we instead adopt a distance of 5\,kpc - the lowest distance estimated for CXOU\,J1714 before it was revised upwards by association with the CTB 37 supernova remnant complex \citep{2012MNRAS.421.2593T} - the transverse velocity estimate drops to $140\pm67$\,kms$^{-1}$, entirely consistent with bound systems. If we change the distance, the inferred absolute magnitude is only affected by a change in distance modulus because the extinction is derived from the column density. Consequently, at 5\,kpc, the source is 2.1 magnitudes fainter, which still places it firmly in the stellar region of parameter space in Figure \ref{fig:cmd}. Although a lower distance is preferred given the VVV proper motion measurement of the NIR source and the typical systemic velocities of X-ray binaries, the large uncertainties mean a reasonable velocity cannot be ruled out even at $d=13.2$\,kpc.

\subsection{Pre-main sequence stars}
Another possibility, if the companion is significantly less massive than the magnetar progenitor, is that it may still be in a pre-main sequence (PMS) period of evolution. The NIR variability of PMS stars is well-established \citep[e.g.][]{carpenter2001, eiroa2002, alves2008}, and several magnetar counterparts are variable. The variations in PMS stars are thought to be due to changes in the disc accretion rate, structure, and clumpy circumstellar dust distributions. Although the time-scales of PMS evolution of low-mass stars are indeed comparable to the entire life-times of massive O stars \citep[tens of Myr;][]{palla1993} -- with the magnetar's life-time adding an insignificant amount of time -- this scenario would invoke the need for a somewhat extreme mass ratio binary. This scenario is disfavoured due to the initial binary parameter distributions \citep{sana2012}, and the preference for these systems to unbind upon primary supernova. However, we know such systems can occur, due to the existence of low-mass X-ray binaries, so the pre-main sequence scenario cannot be ruled out.

\subsection{Comparison with pulsars}
In addition to the population synthesis predictions of Section \ref{sec:bpass}, we can also compare our results to the pulsar population. Seven per cent of pulsars in the ANTF catalogue are listed as having a main sequence companion \citep{2005AJ....129.1993M}\footnote{\url{https://www.atnf.csiro.au/research/pulsar/psrcat/}}. However, this number is highly uncertain due to detection and follow-up biases - some may be obscured by dust, while others have not had deep observations to establish the presence of a companion, or lack thereof. \citet{2021MNRAS.501.1116A} performed a cross-match between the \gaia\ DR3 catalogue and 1534 pulsars. They find that the binary fraction of young pulsars (defined as being un-recycled and with magnetic field strength $>10^{10}$\,G) is $<5$ per cent, under the assumption that all pulsar stellar companions with orbital periods of $<50$\,yr and brighter than 20.5 mag should have been identified. However, some systems with bound companions will manifest as X-ray binaries. We note that the ages of HMXBs, typically tens of Myrs, \citep[inferred from kinematics and cluster associations, e.g.][]{2013ApJ...764..185C} are more typical of (young) pulsars \citep[][]{2005AJ....129.1993M} than magnetars \citep[][]{2014ApJS..212....6O}. This might suggest that more pulsar bound companions will be `missing' (because these systems are instead identified as XRBs) than magnetar bound companions. The lifetimes of bound companions also suggest that they will leave the main sequence on a similar time-scale to the ages of HMXBs, consistent with this scenario. Overall, the population synthesis, pulsar and magnetar estimates for $f_\mathrm{bound}$ are all broadly consistent with ${\sim}$5 per cent.

\subsection{Other constraints and future progress}
The origin of magnetars has been probed in a variety of ways, from investigating their colours/spectral energy distributions (SEDs) as in this work, their proper motions, and associations with recent star formation/supernovae. The results presented here do not require non-core-collapse channels to explain the Galactic population, and this is backed up by previous studies which point towards core-collapse as the dominant channel \citep[e.g.][]{2019MNRAS.487.1426B,2021ApJ...907L..28B}. However, some contribution from other channels cannot be excluded. For example, accretion or merger induced collapse magnetars are predicted on theoretical grounds \citep{1992ApJ...392L...9D,2006MNRAS.368L...1L,2019ApJ...886..110M,2021arXiv210600222G,2021arXiv211012140A}, and are a possible explanation for FRB202001E, which resides in a globular cluster \citep{2021arXiv210511445K,2021arXiv210704059L}. If the majority of magnetars are born in core-collapse events, it is still uncertain whether this is a typical outcome, or if it requires specific pre-collapse conditions \citep[this is largely dependent on how rapidly their magnetic fields decay,][]{2015PASJ...67....9N,2019MNRAS.487.1426B}.

Another avenue to compare magnetars with other neutron stars is their proper motions. Thus far a handful have measured proper motions, \citep[][]{2012ApJ...761...76T,2012ApJ...748L...1D,2013ApJ...772...31T,2020MNRAS.498.3736D}. \cite{2022ApJ...926..121L} also add the proper motion of SGR\,1935 to this distribution, making use of the stability and spatial resolution of \hst . The sample remains small, but they show that there is currently no statistically significant difference between magnetar and pulsar proper motions.

Potential further tests include searching for unbound (runaway) companions. A runaway companion to magnetar CXOU\,J1647 in the young star cluster Westerlund-1 has been claimed \citep[Wd1-5, a blue supergiant,][]{2014A&A...565A..90C}. Due to the carbon rich composition and over-luminous nature of Wd1-5, they infer an evolutionary history in which the (initially slightly less massive) secondary evolves more rapidly following accretion, polluting the primary with carbon rich Wolf-Rayet winds, before going supernova and unbinding the system. Wd1-5 has H${\sim}$8.5, at the upper end of the population synthesis predictions for bound companions (Fig. \ref{fig:bound_H}). We predict that 90 per cent of primary neutron stars should have unbound from their companion following the supernova (47 per cent of core-collapse neutron stars overall, including single stars and mergers). The unbound companions will have a range of velocities, and some will be runaways with velocites of hundreds of kms$^{-1}$. These numbers are less prone to low number statistics and may provide a more robust constraint on magnetar progenitors, but the large number of possible matches means that finding magnetar-secondary pairs is not trivial. Wd1-5 was selected due to its high radial velocity from spectroscopy, but tracing back proper motions may still be a promising route to identifying associations \citep[e.g.][]{2019ApJ...871...92F}. We leave searches for unbound runaway companions, which requires significant additional analyses beyond the scope of this paper, for further studies. However, we qualitatively note that the bound companion search presented here also rules out a fraction of this parameter space, namely unbound companions with small natal kicks.

Another constraint is the masses of the progenitors. For magnetars in a stellar cluster, these can be inferred by ageing the cluster with the red supergiant luminosity distribution. Assuming (i) that the magnetar age is (much) less than the progenitor evolutionary time, and (ii) that the cluster is coeval, the age of the cluster corresponds to a progenitor mass range \citep[e.g.][]{2009ApJ...707..844D}. Progenitor mass estimates range from $>40$\,M$_{\odot}$ \citep[CXOUJ1647, SGR1806][]{2006ApJ...636L..41M,2008MNRAS.386L..23B} down to ${\sim}$15--20M$_{\odot}$ \citep[SGR1900,1E1841][]{,2009ApJ...707..844D,2017ApJ...846...13B}. Supernova remnant associations can also be used to estimate the progenitor mass, based on modelling of the remnant itself. For example, \citet{2019A&A...629A..51Z} use spatially resolved X-ray observations of supernova remnants to infer progenitor masses of less than 20M$_{\odot}$ for magnetars 1E\,1841, SGR\,0526--66 and 1E\,161348--5055. If these progenitor mass distributions can be expanded, comparisons can be made to the overall massive star IMF and pre-SN masses, to see if magnetars represent a biased subset. 

In this work, we have assumed Solar metallicity for the progenitors. If they are biased to lower metallicities, this could be reflected in the progenitor masses and delay times distributions. Quantifying the metallicity of the parent stellar populations is plausible aim, particularly for sources which have cluster associations, e.g. CXOU\,1647 in Westerlund 1 \citep[approximately Solar,][]{2006MNRAS.372.1407C}.

Finally, spectra or SEDs of magnetar counterparts can conclusively determine their nature. They can distinguish stellar from non-stellar, particularly if the SED extends into the $K$-band and beyond, where dusty debris discs start to dominate the spectra \citep{2006Natur.440..772W}. The nature of any stellar companion can also be distinguished. The colours of O-stars and lower-mass pre-main sequence stars differ considerably in the NIR, and also differ from non-thermal/disc models. Characterisation of the NIR SED, as well as a better understanding of the time-scales and amplitude of variability in these sources, would therefore be able to better determine if the emission can be explained by a companion. In principle, neutron star-mass companions could be inferred from radial velocity measurements of bound stellar candidates, but this becomes increasingly difficult at higher mass ratios. Given the locations of magnetars in the Galactic plane, and the associated dusty sight-lines, magnetar spectroscopy and even SED construction is also challenging, but the possibilities will broaden with the advent of {\it JWST} \citep[{\it James Webb Space Telescope},][]{2006SSRv..123..485G}.

\section{Conclusions}\label{sec:conc}
In this paper, we have employed photometry of candidate magnetar counterparts to identify any bound companion stars among the Galactic population. Distance and extinction estimates are used to infer intrinsic colours and absolute magnitudes, which are compared to predictions from binary population synthesis. We find that 5--10 per cent of the Galactic population could plausibly have a bound companion, corresponding to one or two NIR counterparts which appear stellar in nature. One of these is associated with SGR\,0755--2933 and has been identified as a high mass X-ray binary, but it remains unclear whether the accretor in this system is the magnetar, or if this is a chance alignment. The other is the counterpart to CXOU\,J171405.7--381031. These numbers are consistent with binary population synthesis predictions, assuming that the dominant magnetar formation channel is through core-collapse. Further observations are needed to determine if magnetars arise ubiquitously from core-collapse events or a biased subset of this population, and to what extent alternative progenitor channels contribute.

\section*{Acknowledgements}
AAC is supported by the Radboud Excellence Initiative. AJL has
received funding from the European Research Council (ERC) under
the European Union’s Seventh Framework Programme (FP7-2007-
2013) (Grant agreement No. 725246). PJG acknowledges support from the NRF SARChI program under grant number 111692. JDL acknowledges support from a UK Research and Innovation Fellowship (MR/T020784/1). CK and ASF acknowledge support for this research, provided by NASA through a grant from the Space Telescope Science Institute, which is operated by the Association of Universities for Research in Astronomy, Inc. We thank Rens Waters and Gijs Nelemans for useful discussions.

This work made use of v2.2.1 of the Binary Population and Spectral Synthesis (BPASS) models as described in \citet{2017PASA...34...58E} and \citet{2018MNRAS.479...75S}. This work has made use of {\sc ipython} \citep{2007CSE.....9c..21P}, {\sc numpy} \citep{2020arXiv200610256H}, {\sc scipy} \citep{2020NatMe..17..261V}; {\sc matplotlib} \citep{2007CSE.....9...90H}, Seaborn packages \citep{Waskom2021}, {\sc astropy},\footnote{https://www.astropy.org} a community-developed core Python package for Astronomy \citep{astropy:2013, astropy:2018}, and the python package for BPASS, HOKI \citep{2020JOSS....5.1987S}. We have also made use of the python modules {\sc statsmodels} \citep{seabold2010statsmodels}, {\sc dustmaps} \citep{2018JOSS....3..695M}, {\sc extinction} \citep{barbary_kyle_2016_804967} and {\sc photutils}, an Astropy package for detection and photometry of astronomical sources \citep{larry_bradley_2021_4624996}. This research has made use of the SVO Filter Profile Service (\url{http://svo2.cab.inta-csic.es/theory/fps/}) supported from the Spanish MINECO through grant AYA2017-84089 \citep{2012ivoa.rept.1015R,2020sea..confE.182R}. 

This publication makes use of data products from the Two Micron All Sky Survey, which is a joint project of the University of Massachusetts and the Infrared Processing and Analysis Center/California Institute of Technology, funded by the National Aeronautics and Space Administration and the National Science Foundation. This work has made use of data from the European Space Agency (ESA) mission {\it Gaia} (\url{https://www.cosmos.esa.int/gaia}), processed by the {\it Gaia} Data Processing and Analysis Consortium (DPAC,
\url{https://www.cosmos.esa.int/web/gaia/dpac/consortium}). Funding for the DPAC has been provided by national institutions, in particular the institutions participating in the {\it Gaia} Multilateral Agreement.

Finally, we thank the referee for their constructive comments on this manuscript.

\section*{Data Availability}
Based on observations made with the NASA/ESA Hubble Space Telescope, obtained from the data archive at the Space Telescope Science Institute. STScI is operated by the Association of Universities for Research in Astronomy, Inc. under NASA contract NAS 5--26555. These observations are associated with programs 14805, 15348 and 16019 (Levan). The scientific results reported in this article are based on data obtained from the Chandra Data Archive.




\bibliographystyle{mnras}
\bibliography{magnetarbinaries} 




\appendix

\section{Colour-magnitude diagrams at other ages}\label{apx:cmdages}
\begin{figure*}
	\includegraphics[width=0.49\textwidth]{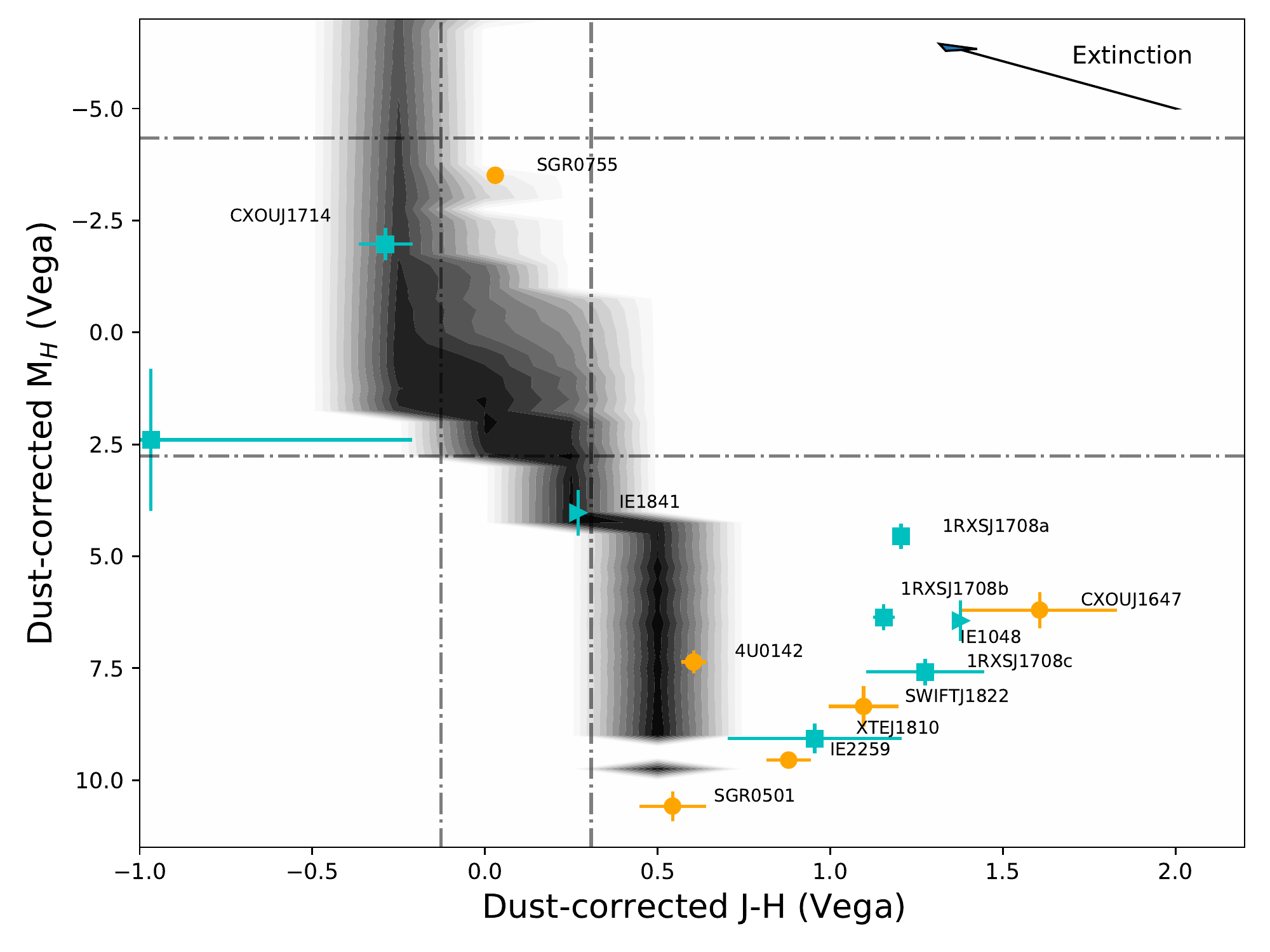}
	\includegraphics[width=0.49\textwidth]{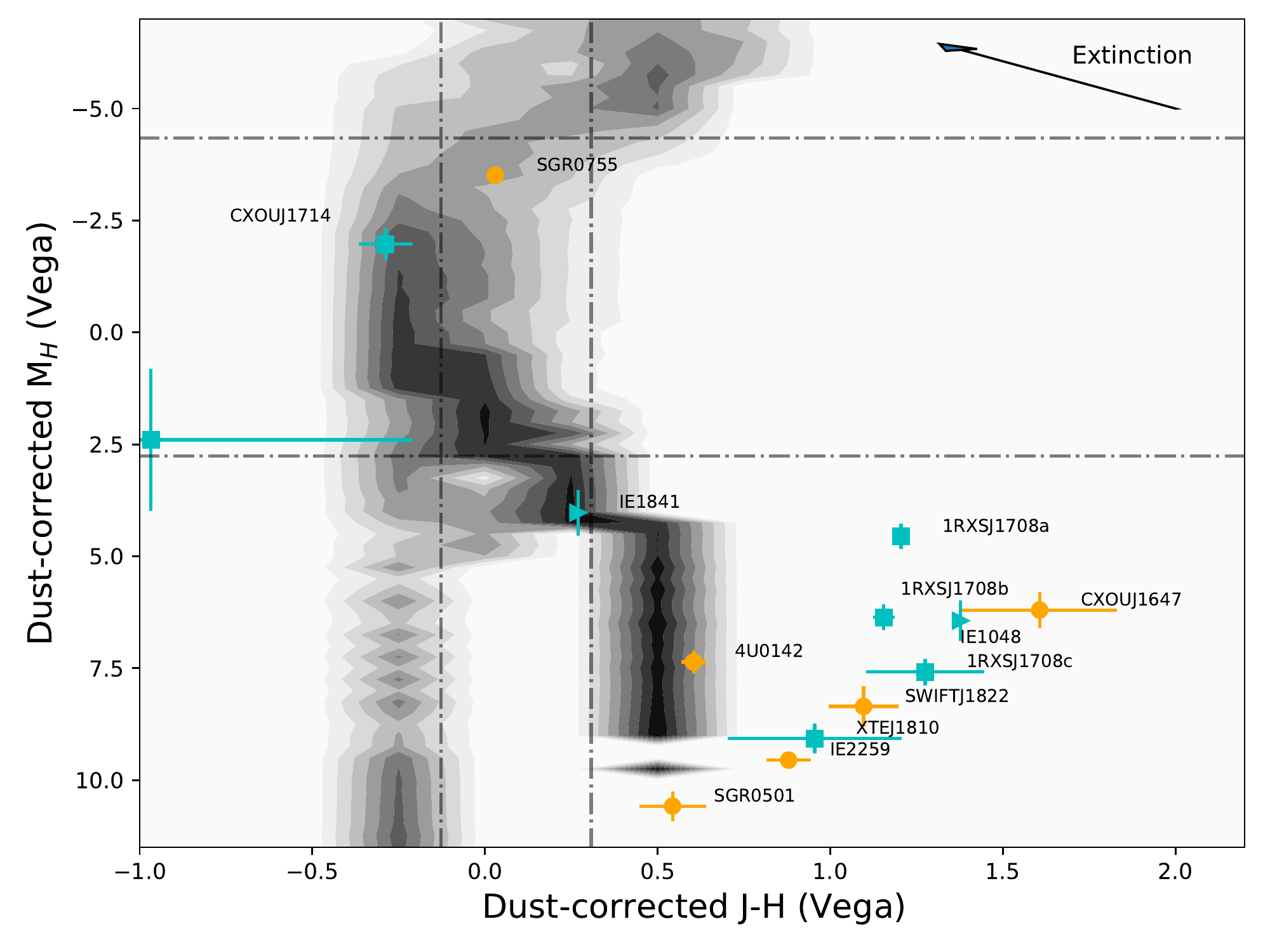}
    \caption{Left: a BPASS NIR colour-magnitude diagram made with HOKI, as in Figure \ref{fig:cmd}, at $10^{6}$\,yr rather than $10^{7}$\,yr. Right: the same plot at $10^{8}$\,yr. The age range of 1 to 100\,Myr represents the expected lifetimes of supernova progenitors; the stellar populations shown therefore represent the possible $J$ and $H$ photometric properties of all stellar objects at the time of primary star supernovae, without restricting ourselves specifically to the predicted properties of bound companions.}
    \label{fig:apx}
\end{figure*}

In this appendix, we show colour-magnitude diagrams at two other ages, $10^{6}$\,yr and $10^{8}$\,yr, covering the majority of the age range where core-collapse supernovae are expected (Figure \ref{fig:apx}). Along with Figure \ref{fig:cmd}, we therefore show the range of stellar population colours and magnitudes at the expected age of bound companions to recently born core-collapse neutron stars, agnostic to any prior assumptions about the expected properties of bound companions. 

\bsp	
\label{lastpage}
\end{document}